# IMPETUS: Consistent SPH calculations of 3D spherical Bondi accretion onto a black hole

J. M. Ramírez-Velasquez,[1,2,3]⋆ L. Di G. Sigalotti,[4] R. Gabbasov,[5] F. Cruz,[4] J. Klapp[2,6]

[1]*Centro de Física, Instituto Venezolano de Investigaciones Científicas (IVIC), Apartado Postal 20632, Caracas 1020A, Venezuela*

[2]ABACUS-*Centro de Matemática Aplicada y Cómputo de Alto Rendimiento, Departamento de Matemáticas, Centro de Investigación y de Estudios Avanzados (Cinvestav-IPN), Carretera México-Toluca km. 38.5, La Marquesa, 52740 Ocoyoacac, Estado de México,Mexico*

[3]*ESPOL Polytechnic University, Escuela Superior Politécnica del Litoral (ESPOL), Physics Department, FCNM, Campus Gustavo Galindo km. 30.5, Vía Perimetral, P.O. Box 09-01-5863, Guayaquil, Ecuador*

[4]*Área de Física de Procesos Irreversibles, Departamento de Ciencias Básicas, Universidad Autónoma Metropolitana-Azcapotzalco (UAM-A), Av. San Pablo 180, 02200 Ciudad de México, Mexico*

[5]*Instituto de Ciencias Básicas e Ingenierías, Universidad Autónoma del Estado de Hidalgo (UAEH), Ciudad Universitaria, Carretera Pachuca-Tulancingo km. 4.5 S/N, Colonia Carboneras, Mineral de la Reforma, C.P. 42184, Hidalgo, Mexico*

[6]*Departamento de Física, Instituto Nacional de Investigaciones Nucleares (ININ), Carretera México-Toluca km. 36.5, La Marquesa, 52750 Ocoyoacac, Estado de México, Mexico*

**ABSTRACT**

We present three-dimensional calculations of spherically symmetric Bondi accretion onto a stationary supermassive black hole (SMBH) of mass $10^8\ M_\odot$ within a radial range of $0.02-10$ pc, using a modified version of the smoothed particle hydrodynamics (SPH) GADGET-2 code, which ensures approximate first-order consistency (i.e., second-order accuracy) for the particle approximation. First-order consistency is restored by allowing the number of neighbours, $n_{\rm neigh}$, and the smoothing length, $h$, to vary with the total number of particles, $N$, such that the asymptotic limits $n_{\rm neigh}\to\infty$ and $h\to 0$ hold as $N\to\infty$. The ability of the method to reproduce the isothermal ($\gamma=1$) and adiabatic ($\gamma=5/3$) Bondi accretion is investigated with increased spatial resolution. In particular, for the isothermal models the numerical radial profiles closely match the Bondi solution, except near the accretor, where the density and radial velocity are slightly underestimated. However, as $n_{\rm neigh}$ is increased and $h$ is decreased, the calculations approach first-order consistency and the deviations from the Bondi solution decrease. The density and radial velocity profiles for the adiabatic models are qualitatively similar to those for the isothermal Bondi accretion. Steady-state Bondi accretion is reproduced by the highly resolved consistent models with a percent relative error of $\lesssim 1\%$ for $\gamma=1$ and $\sim 9\%$ for $\gamma=5/3$, with the adiabatic accretion taking longer than the isothermal case to reach steady flow. The performance of the method is assessed by comparing the results with those obtained using the standard GADGET-2 and the GIZMO codes.

## 1 INTRODUCTION

There is irrefutable observational evidence that active galaxies harbour supermassive black holes (SMBHs) at their centres, with masses ranging from $\sim 10^6$ to $\sim 10^9$ $M_\odot$. Since SMBHs grow by accretion of the surrounding gas and since feedback processes from them are important in regulating the growth and evolution of the host galaxies, the study of the central galactic parsec regions has become a major research field in modern astrophysics (Salpeter 1964; Lynden-Bell 1969; Blandford 1976; Ozernoi & Reinhardt 1978; Balick & Heckman 1982; Rees 1984; Fabian & Crawford 1990; Kormendy & Richstone 1995; Kauffmann & Haehnelt 2000; Ferrarese & Ford 2005). The gravitational influence of SMBHs is known to be particularly strong within the Bondi radius

$$R_{\rm B}\approx 3\left(\frac{M_{\rm BH}}{10^8\ M_\odot}\right)\left(\frac{T_\infty}{10^7\ {\rm K}}\right)^{-1}\ {\rm pc}, \tag{1}$$

⋆ E-mail: josem.ramirez@gmail.com

where $M_{\rm BH}$ is the mass of the SMBH and $T_\infty$ is the temperature of the environment surrounding it. In its seminal work, Bondi (1952) established that, at least for spherically symmetric flows, the surrounding medium will definitely fall into an accretion process if it is close to $R_{\rm B}$.

Studies of the physical properties of accretion onto compact objects have been mainly motivated by the observables that we measure from the light reaching the Earth. For example, each band of the electromagnetic spectrum from radio to $\gamma$-rays has allowed the study of several phenomena that can now be simulated in a computer, from huge radio lobes emission produced by parsec-scale jets in the centre of active galactic nuclei (AGNs) to the X-ray emission from sub-parsec scales during the accretion onto a SMBH in the centre of galaxies. Observational evidence for X-ray emitting stellar-mass and intermediate-mass black holes, which are thought to be born during the core collapse of massive stars and with estimated masses ranging from $\sim 4$ to 14 $M_\odot$ or more (Orosz 2003), have also become compelling (Miller & Colbert 2004; King & Lasota 2016). Moreover, direct signature of Bondi accretion onto a SMBH of mass $2\times10^9$ $M_\odot$ has been revealed by a *Chandra* observation of the hot gas in NGC 3115 (Wong et al. 2011). However, it is not clear whether in the centre of galaxies Bondi accretion rates are the dominant components of the inflow material. Also, emission of hard X-ray photons as observed in real systems with the new generation of X-ray telescopes may well be originating in these hot and dense regions (Braito et al. 2007; Farrell et al. 2009; Dewangan et al. 2013; Russell et al. 2015; Luangtip et al. 2016). Attempts to construct the energy spectra of Ultra-Luminous X-ray sources from general relativistic radiation magnetohydrodynamics simulations of super-Eddington accretion onto a 10 $M_\odot$ black hole have recently been reported by Narayan et al. (2017). These extreme phenomena have had a profound impact on our present knowledge of the relationship between the growth of the SMBH and the evolution of its host galaxy. Yet, a number of key issues remain poorly understood as, for example, the interpretation and decomposition of the energy spectra of accreting black holes, which are much debated today.

Astrophysical processes involving radiative energy transfer are calculated by the balance between compressional heating and adiabatic cooling (Blondin et al. 1990; Taam et al. 1991). Analytical prescriptions for the heating and cooling rates in complex environments are only possible under certain limits. However, the increasing computer power available today is allowing to model complex astrophysical scenarios efficiently and at a relatively low cost, including the dynamical update of the microphysics and chemistry. For instance, thermal instabilities, non-axysimmetric rotating objects, the relationships between ionization, molecular states, level populations, and kinetic temperatures of low density environments are some of the ingredients that have no analytical counterparts and that can be calculated with highly efficient numerical algorithms. A particularly interesting question is related to the accretion of angular momentum, when the flow contains a density or velocity gradient perpendicular to the flow direction (Davies & Pringle 1980; Ruffert & Anzer 1995; Ruffert 1997, 1999). In particular, Proga & Begelman (2003) performed two-dimensional (2D) axisymmetric simulations of the accretion of slowly rotating gas onto a black hole. They found that for low angular momentum, the flow is Bondi-like, while for intermediate total angular momentum, the gas with higher angular momentum does not accrete and forms a dense torus about the black hole and, as a result, the accretion rate is substantially below the Bondi rate. For even higher total angular momentum, a dense torus coupled to a centrifugal barrier sets in, which further reduces the accretion rate. Proga & Begelman (2003) have argued that rotation may explain the difference between accretion rates estimated from observations and those predicted by the Bondi flow. Moreover, Perna et al. (2003) suggested that combining rotation and magnetic fields results in accretion rates onto isolated neutron stars that are much lower than those estimated using the Bondi formula. Later on, Krumholz et al. (2005) considered the three-dimensional (3D) accretion of gas with constant vorticity and found that the resulting flow field is highly non-axisymmetric and time-dependent and that even a small amount of vorticity can substantially change the accretion rate. Considerable progress has been made over the past twenty years in developing advanced numerical models to understand the physics of accretion onto compact objects. A thorough review of Bondi accretion theory and related earlier numerical simulations is given by Edgar (2004).

A further important aspect of the problem that prompted an intense theoretical and numerical investigation is the instability of the Bondi accretion flow. Since the analysis performed by Cowie (1977), who showed that the Bondi model is unstable against short-wavelength density perturbations, and the extension of Cowie's (1977) work by Soker (1990) to include tangential oscillations (corresponding to the so-called 'flip-flop' instability), the instability has been observed in many numerical simulations and analyzed through known physical mechanisms and possible numerical artefacts. An overview of existing simulations and proposed instability mechanisms is given by Foglizzo et al. (2005). Whereas the instability of the 3D Bondi accretion flow has not been completely clarified yet, more recent work of 2D planar accretion showed that the flip-flop instability is a true overstability rather than a numerical artefact (Blondin & Pope 2009), while in three dimensions the flow remained remarkably axisymmetric and displayed a relatively stable bow shock and no flip-flop behaviour (Blondin & Raymer 2012). In this case, a smaller accretor was seen to produce a less stable flow pattern.

It is well-known that the accretion onto compact objects may influence the nearby ambient around SMBHs in the centre of galaxies (e.g., Salpeter 1964; Fabian 1999; Barai 2008; Germain et al. 2009). Together with the outflow phenomena, it is believed to play a major role in the feedback processes invoked by modern cosmological models (i.e., $\Lambda$-Cold Dark Matter) to explain the possible relationship between the SMBH and its host galaxy (e.g., Magorrian et al. 1998; Gebhardt et al. 2000) as well as in the self-regulating growth of the SMBH. The problem of accretion onto a SMBH has been studied via hydrodynamical simulations (e.g., Ciotti & Ostriker 2001; Di Matteo et al. 2005; Li et al. 2007; Ostriker et al. 2010; Novak et al. 2011). In numerical studies of galaxy formation, spatial resolution permits resolving scales from the kpc to the pc, while the sub-parsec scales of the Bondi radius are not resolved.

This is why a prescribed sub-grid physics is employed to solve this lack of resolution. With sufficiently high X-ray luminosities, the falling material will have the correct opacity, developing outflows that originate at sub-parsec scales. Therefore, calculations of the processes involving accretion onto SMBH have become of primary importance (e.g., Proga et al. 2000; Proga 2000, 2003; Proga & Kallman 2004; Proga 2007; Ostriker et al. 2010). Numerical calculations of the accretion of matter onto SMBHs, including the radiative-outflow component, have been mostly performed using Eulerian finite-difference methods (Mościbrodzka & Proga 2013) (see also the overviews by Edgar (2004) and Foglizzo et al. (2005) and references therein for earlier work), while only a few calculations have been reported using Smoothed Particle Hydrodynamics (SPH) techniques (Barai 2008; Barai et al. 2011, 2012; Nagamine et al. 2012), where results for the accretion rates, outflow rates, thermal instabilities, and impact of the thermal, mechanical, and X-ray feedbacks have been obtained for evolutions up to 5 Myr and scales from 0.1 to 200 pc.

Among the several existing codes for the calculation of accretion onto SMBHs, ZEUS (Hayes et al. 2006) and GADGET-2 (Springel 2005) along with the more recently released version GADGET-3 and the public version of the GIZMO code (Hopkins 2015) have been the most widely used *state-of-the-art* gravitohydrodynamics codes. Numerical solutions to the equations of gravitohydrodynamics in full 3D have been often addressed using SPH methods, where fluid elements are represented by a discrete set of particles that characterize the flow attributes. The idea of "smoothing" the properties of a particle from those of its own neighbours to get insight into its properties, has become increasingly popular in astrophysics and cosmology to simulate a variety of complex problems, ranging from protoplanetary disks (Moeckel & Throop 2009) to star formation processes (Arreaga-García et al. 2007; Commerçon et al. 2008; Bürzle et al. 2011; Riaz et al. 2014; Gabbasov et al. 2017) to accretion onto supermassive black holes (Barai et al. 2011, 2012) to the formation of first stars and galaxies (Ricotti 2007) to cosmic structure formation (Springel 2005; Springel et al. 2005; Riebe et al. 2013).

In this paper, we perform small-scale simulations of 3D spherical gas accretion onto a stationary SMBH using a modified version of the GADGET-2 code (Springel 2005), which ensures approximate first-order SPH consistency (or second-order accuracy) even in the presence of varying smoothing lengths (Gabbasov et al. 2017). The rationale of using a consistent SPH scheme over other more traditional methods is that SPH does not suffer from grid alignment and therefore it is worth assessing its ability to accurately reproduce the Bondi mass accretion at sub-parsec scales. Compared to previous SPH calculations of gas accretion onto a SMBH, we take a different approach in which the number of neighbours, $n_{\rm neigh}$, is allowed to vary with the total number of particles, $N$, such that the combined limit $N\to\infty$, $h\to 0$, and $n_{\rm neigh}\to\infty$ with $n_{\rm neigh}/N\to 0$ for full SPH consistency is achieved as the domain resolution is increased (Rasio 2000; Zhu et al. 2015; Sigalotti et al. 2016a), where $h$ is the smoothing length. To do so, we implement the parameterizations devised by Zhu et al. (2015), where $n_{\rm neigh}$ and $h$ are allowed to vary with $N$ so that the above joint limit is satisfied as $N\to\infty$. This realization is consistent with the error analysis of the SPH continuity and momentum equations performed by Read et al. (2010), who found that SPH consistency is lost due to zeroth-order errors that would persist when working with a fixed number of neighbours even though $N\to\infty$ and $h\to 0$. The improved SPH consistency of our modified code is measured by comparing the quality of the particle consistency constraints with that resulting from identical SPH models using the standard GADGET-2 code with fixed $n_{\rm neigh}(=64)$. Due to the lack of numerical solutions for the 3D spherical Bondi accretion, the results are validated by direct comparison with identical models using the particle-based finite-volume GIZMO code (Hopkins 2015).

Our long-term goal is therefore to develop a consistent SPH model of AGN feedback (Kurosawa et al. 2009), where future work will account for the radiative-outflow component and improved microphysics (Ramírez-Velasquez et al. 2016). This paper is organized as follows. Section 2 deals with the issue of consistency in SPH. The SPH solver and the model set-up are briefly described in Section 3, while the results are presented in Section 4. Section 5 contains a discussion on the instability of the Bondi flow and Section 6 summarizes the conclusions.

## 2 SPH CONSISTENCY CONSTRAINTS

In standard SPH the smoothed estimate of a function, say $f(\mathbf{r})$, is defined by the integral interpolant

$$\langle f(\mathbf{r})\rangle = \int_{\Omega} f(\mathbf{r}')W(|\mathbf{r}-\mathbf{r}'|,h)d^3\mathbf{r}' \tag{2}$$

where $\mathbf{\Omega}$ is the spatial domain, $W$ is the interpolation kernel approximating the Dirac-$\delta$ distribution, and $h$ is the smoothing length or width of the kernel. The kernel interpolation function $W$ must: (a) approach the Dirac-$\delta$ distribution as $h\to 0$, (b) satisfy the normalization condition

$$M_0 = \int_{\Omega} W(|\mathbf{r}-\mathbf{r}'|,h)d^3\mathbf{r}' = 1, \tag{3}$$

and (c) be a positive-definite, symmetric, and monotonically decreasing function centred on the particle position $\mathbf{r}$. In addition, the kernel is chosen to have compact support so that $W(|\mathbf{r}-\mathbf{r}'|,h)=0$ beyond a finite extent $|\mathbf{r}-\mathbf{r}'|>kh$, where $kh$ is the radius of the kernel support and $k$ is some number usually $\leq 2$. Also, replacing $|\mathbf{r}-\mathbf{r}'|$ by $h|\mathbf{r}-\mathbf{r}'|$ in the form of $W$ yields the relation

$$W(h|\mathbf{r}-\mathbf{r}'|,h) = \frac{1}{h^3}W(|\mathbf{r}-\mathbf{r}'|,1), \tag{4}$$

which holds for any suitable SPH kernel function (Sigalotti et al. 2016a). The same is also true for the gradient of the kernel.

After expanding in Taylor series $f(\mathbf{r}')$ around $\mathbf{r}'=\mathbf{r}$, making the change of variables $\mathbf{r}\to h\mathbf{r}$ and $\mathbf{r}'\to h\mathbf{r}'$, using relation (4), and plugging the result into Eq. (2) yields the family of constraints for the moments of the kernel

$$\mathbf{M}_l = \int_{\Omega} (\mathbf{r}'-\mathbf{r})^l W(|\mathbf{r}-\mathbf{r}'|,1)d^3\mathbf{r}' = \mathbf{0}^{(l)}, \quad \text{for} \quad l=1,2,...,m, \tag{5}$$

that must be satisfied for exact interpolation of a polynomial to order $m+1$ with the use of Eq. (2), where $(\mathbf{r}'-\mathbf{r})^l$ is a tensor of rank $l$ and $\mathbf{0}^{(l)}$ denotes the zero tensor of rank $l$. Satisfaction of the integral relations (3) and (5) implies $C^m$-consistency for the kernel approximation. Note that the zeroth moment of the kernel is $M_0 = 1$, which is just the normalization condition (3).

Similar consistency constraints can be derived for the estimate of the gradient

$$\langle \nabla f(\mathbf{r})\rangle = \int_\Omega f(\mathbf{r}')\nabla W(|\mathbf{r}-\mathbf{r}'|,h)d^3\mathbf{r}', \quad (6)$$

as

$$\mathbf{M}'_0 = \int_\Omega \nabla W(|\mathbf{r}-\mathbf{r}'|,1)d^3\mathbf{r}' = \mathbf{0}, \quad (7)$$

$$\mathbf{M}'_1 = \int_\Omega (\mathbf{r}'-\mathbf{r})\nabla W(|\mathbf{r}-\mathbf{r}'|,1)d^3\mathbf{r}' = \mathbf{I}, \quad (8)$$

$$\mathbf{M}'_l = \int_\Omega (\mathbf{r}'-\mathbf{r})^l\nabla W(|\mathbf{r}-\mathbf{r}'|,1)d^3\mathbf{r}' = \mathbf{0}^{(l+1)}, \quad \text{for} \quad l = 2,3,...,m, \quad (9)$$

where $\mathbf{0}$ is the null vector, $\mathbf{I}$ is the unit tensor, $\mathbf{0}^{(l+1)}$ is the zero tensor of rank $l+1$, and the nabla operator is taken with respect to coordinates $\mathbf{r}$. However, after a simple algebra the moments of the gradient (7)-(9) can be written in terms of the moments of the kernel as

$$\mathbf{M}'_1 = M_0\mathbf{I}, \quad (10)$$

$$\mathbf{M}'_l = l\mathbf{M}_{l-1}\mathbf{I}, \quad \text{for} \quad l = 2,3,...,m, \quad (11)$$

where the order of the tensor product $\mathbf{M}_{l-1}\mathbf{I}$ does not lead to ambiguities because $\mathbf{M}_{l-1}$ and $\mathbf{I}$ are both symmetric tensors. In practice, from relations (3), (5), and (7)-(9) we may see that $C^0$- and $C^1$-consistencies are automatically satisfied by the kernel approximation, while $C^2$-consistency is not easily achieved unless the kernel function approaches the Dirac-$\delta$ distribution in the limit $h \to 0$. This can be seen by noting that in general

$$\mathbf{M}_2 = \langle \mathbf{r}^2\rangle - \langle \mathbf{r}\rangle\langle \mathbf{r}\rangle \neq \mathbf{0}^{(2)}, \quad (12)$$

and so according to relation (11) $\mathbf{M}'_3 = 3\mathbf{M}_2\mathbf{I} \neq \mathbf{0}^{(4)}$. Hence, the second moment of the kernel is just the variance of the particle position vector $\mathbf{r}$ and is a measure of the spread of the particle positions relative to the mean. While these results are independent of the form of the kernel, the non-vanishing moments $\mathbf{M}_2$ and $\mathbf{M}'_3$ imply truncation errors of $O(h^2)$ for both the kernel approximations of the function and its gradient.

It is well-known that the consistency constraints (3), (5), and (7)-(9) are in general not satisfied by the particle approximation, i.e., when the integrals are replaced by sums over a discrete set of particles within the kernel support. In particular, loss of $C^0$-consistency for the particle approximation of the function and its gradient occurs because

$$M_{0,a} = \sum_{b=1}^{n_{\rm neigh}} \frac{m_b}{\rho_b} W_{ab} \neq 1, \quad (13)$$

and

$$\mathbf{M}'_{1,a} = \sum_{b=1}^{n_{\rm neigh}} \frac{m_b}{\rho_b} \mathbf{r}_{ba}\nabla_a W_{ab} \neq \mathbf{I}, \quad (14)$$

where the subscripts $a$ and $b$ are particle labels, $W_{ab} = W(|\mathbf{r}_a - \mathbf{r}_b|,1)$, $\mathbf{r}_{ba} = \mathbf{r}_b - \mathbf{r}_a$, $m_b$ is the mass of neighbour particle $b$, and $\rho_b$ is its density. If, on the other hand, $C^0$ particle consistency is restored, then $C^1$ particle consistency is also ensured because of the symmetry of the kernel. In fact, recent protostellar collapse calculations have shown that independently of the spatial resolution employed the discrete consistency relations $\mathbf{M}_{1,a} = \mathbf{0}$, $\mathbf{M}'_{0,a} = \mathbf{0}$, and $\mathbf{M}'_{2,a} = \mathbf{0}^{(3)}$ were reproduced numerically during the full evolutions (Gabbasov et al. 2017). Evidently, the quality of the sums (13) and (14) will depend only on the number of neighbours, $n_{\rm neigh}$, and their actual spatial distribution within the kernel support. Using a simple linear analysis on one-dimensional sound wave propagation, Rasio (2000) found that full particle consistency in SPH is possible only if the joint limit $N \to \infty$, $h \to 0$, and $n_{\rm neigh} \to \infty$ holds with $n_{\rm neigh}/N \to 0$, where $N$ is the total number of particles filling the computational domain. This result is consistent with the error analysis of the SPH continuity and momentum equations carried out by Read et al. (2010), who found that working with finite values of $n_{\rm neigh}$ induces zeroth-order errors that would scale with $n_{\rm neigh}$ as $\sim n_{\rm neigh}^{-1}$ even when $N \to \infty$ and $h \to 0$. Based on a balance between the kernel and particle approximation errors, Zhu et al. (2015) derived the scaling relations $h \propto N^{-1/\beta}$ and $n_{\rm neigh} \propto N^{1-3/\beta}$ such that $h \propto n_{\rm neigh}^{1/(3-\beta)}$, with $5 \leq \beta \leq 7$. In particular, these parameterizations comply with the joint limit $N \to \infty$, $h \to 0$, and $n_{\rm neigh} \to \infty$ for full SPH consistency as $N$ is increased.

A recent analysis has demonstrated that using the above scalings $C^0$-consistency is restored for both the function and its derivatives when using standard SPH, with the numerical solution becoming insensitive to the degree of particle disorder (Sigalotti et al. 2016a). Moreover, a novel approach using the Poisson summation formula has shown that the discretization error in SPH goes as (Sigalotti et al. 2016b)

$$E \sim \frac{a_0}{n_{\rm neigh}} + \frac{a_1 h}{n_{\rm neigh}} + \left(\frac{a_2}{n_{\rm neigh}} + a_2^{(\rm K)}\right)h^2, \quad (15)$$

where the term with $a_2^{(\rm K)}$ is the contribution from the continuous kernel approximation. For large $n_{\rm neigh}$ the error is dominated by the term $a_2^{(\rm K)}h^2$, while for small $n_{\rm neigh}$ and $h$ the error is given by the zeroth-order term $a_0/n_{\rm neigh}$. This shows that in the limit $n_{\rm neigh} \to \infty$ the particle discretization error vanishes and the particle estimate of a function tends to its kernel estimate, thereby restoring full $C^1$-consistency (or second-order accuracy). With this in mind, we study the consistency of our SPH scheme using as a benchmark test the 3D spherical Bondi accretion onto a stationary SMBH by scaling the smoothing length and the number of neighbours with the allowed resolution according to Zhu et al. (2015) parameterizations. The degree of consistency is judged by the quality of the moments $M_{0,a}$ and $\mathbf{M}'_{1,a}$ defined by Eqs. (13) and (14), respectively, for increasing spatial resolution. The Bondi accretion problem is particularly useful for testing the consistency of the method because it admits an exact solution.

## 3 NUMERICAL METHODS AND MODEL SET-UP

### 3.1 SPH formulation

We perform two separate sequences of calculations for the isothermal and abiabatic spherical Bondi accretion onto a SMBH. One sequence uses the standard GADGET-2 code described by Springel (2005), where the SPH interpolation is performed using the standard cubic $B$-spline kernel with fixed $n_{\rm neigh}(=64)$ independently of $N$, while the other sequence is calculated using a modified version of the code, where $n_{\rm neigh}$ is allowed to vary with increasing $N$ according to the consistency parameterizations suggested by Zhu et al. (2015).

In our modified version of the code the cubic $B$-spline kernel is replaced by a Wendland C$^4$ function (Wendland 1995; Dehnen & Aly 2012)

$$W(q,h) = \frac{495}{32\pi h^3}\begin{cases}(1-q)^6\left(1+6q+\frac{35}{3}q^2\right), & q\leq 1,\\ 0, & \text{otherwise,}\end{cases} \tag{16}$$

where $q = |\mathbf{r}-\mathbf{r}'|/h$. Unlike most commonly used SPH kernels, Wendland functions have the ability to support arbitrarily large numbers of neighbours and suppress unstable behaviour due to close pairing of particles. Even in highly dynamical tests, they exclude particle motion on a sub-scale resolution and maintain the particles more regularly distributed compared to traditional kernels (Rosswog 2015). Zhu et al. (2015) demonstrated that for truly random data points the standard deviation of the density distribution significantly deviates from the expected trend $n_{\rm neigh}^{-0.5}$ when $q=0$ (i.e., when the particle self-contribution is included). This produces an overestimate of the density where the particles are randomly distributed. The bias towards $q=0$ at small $n_{\rm neigh}$ can be greatly reduced by subtracting the self-contribution term. However, they also showed that for a glass-like (or quasi-regular) distribution of particles there is no need to exclude the self-contribution term since the standard deviation of the density distribution matches the $n_{\rm neigh}^{-1}$ trend for all values of $n_{\rm neigh}$. In real SPH applications, the distances between neighbour pairs tend to equilibrate due to pressure forces, which makes the interpolation errors much smaller and the particle distribution more ordered than for a random distribution. In this sense, a random configuration represents an extreme case for SPH simulations. In contrast, a glass configuration represents the other extreme case where the particles are quasi-regularly distributed and almost force-free. While realistic applications fall between these two extremes, we expect the effects of the particle self-contribution to be negligible at large $n_{\rm neigh}$ even in zones where the particle distribution is highly disordered.

A further important improvement concerns the artificial viscosity switch, which is implemented using the method proposed by Hu et al. (2014). In this method the artificial viscosity acceleration term entering the momentum equation has the same form used in GADGET-2 but a different formulation for the artificial viscosity coefficient $\alpha_a$ of a particle. The same principles of the method developed by Cullen & Dehnen (2010) hold, except that now the method is equipped with a stronger limiter that applies the same weight to the velocity divergence, $\nabla\cdot\mathbf{v}$, and vorticity, $\nabla\times\mathbf{v}$. This has the benefits of suppressing excessive viscous dissipation in subsonically convergent flows and ensuring a rapid rise of $\alpha_a$ to a prescribed maximum value ($\alpha_{\rm max}=0.1$) when the flow becomes supersonic. This is a particularly important property in accretion simulations where unphysical dissipation of local velocity differences away from shocks is strongly reduced.

### 3.2 The Bondi problem

The spherically-symmetric accretion onto a central stationary object was examined by Bondi (1952) for the case when the accreting gas is at rest at infinity and free of self-gravity. Far from the central object the gas is assumed to have uniform density $\rho_\infty$ and pressure $p_\infty$. The gas accretion is steady and the central object is assumed to be a SMBH of mass $M_{\rm BH}$. The increase of mass of the SMBH is ignored and, since the gas self-gravity is neglected, the only external force acting on the gas is the gravitational force due to the SMBH. The equation of state, relating the pressure and density, is $p\propto\rho^\gamma$, with a polytropic index $1\leq\gamma\leq 5/3$.

A solution for the gas radial velocity $v$ and density $\rho$ as a function of radius $r$ can be obtained by solving the continuity equation

$$\dot{M} = -4\pi r^2\rho v \equiv \text{constant}, \tag{17}$$

coupled to the Bernoulli integral

$$\frac{v^2}{2} + \left(\frac{\gamma}{\gamma-1}\right)\frac{p_\infty}{\rho_\infty}\left[\left(\frac{\rho}{\rho_\infty}\right)^{\gamma-1}-1\right] = \frac{GM_{\rm BH}}{r}, \tag{18}$$

where the term on the right-hand side of Eq. (18) is the Newtonian gravitational potential of the central SMBH. Among the family of possible solutions to Eqs. (17) and (18), the relevant one for accretion flows is that for which the flow is subsonic at large radii $r>R_{\rm s}$ and becomes supersonic close to the SMBH, i.e., when $r<R_{\rm s}$, where $R_{\rm s}$ is the sonic point. The transonic accretion rate for this solution is

$$\dot{M}_{\rm B} = 4\pi\lambda_{\rm c}\frac{(GM_{\rm BH})^2}{c_{\rm s,\infty}^3}\rho_\infty, \tag{19}$$

where $c_{\rm s,\infty} = (\gamma p_\infty/\rho_\infty)^{1/2}$ is the sound speed at infinity, the ratio $p_\infty/\rho_\infty$ for a gas at temperature $T_\infty$ and molecular weight $\mu$ is $p_\infty/\rho_\infty = k_{\rm B}T_\infty/(m_{\rm p}\mu)$, where $k_{\rm B}$ is the Boltzmann constant, $m_{\rm p}$ is the proton mass, and $\mu=0.63$. The parameter $\lambda_{\rm c}$ is given by

$$\lambda_{\rm c} = \left(\frac{1}{2}\right)^{(\gamma+1)/(2\gamma-2)}\left(\frac{5-3\gamma}{4}\right)^{(3\gamma-5)/(2\gamma-2)}, \tag{20}$$

and the sonic radius is defined by

$$R_{\rm s} = \left(\frac{5-3\gamma}{4}\right)R_{\rm B}, \tag{21}$$

where

$$R_{\rm B} = \frac{GM_{\rm BH}}{c_{\rm s,\infty}^2}. \tag{22}$$

is the Bondi radius. The sound crossing time from $R_{\rm B}$ to the centre is defined as $t_{\rm B} = R_{\rm B}/c_{\rm s,\infty}$. When the adiabatic index is $\gamma=1$, the sonic radius is $R_{\rm s} = R_{\rm B}/2$, while for $\gamma=5/3$ it vanishes (i.e., $R_{\rm s}=0$). However, for these special values the parameter $\lambda_{\rm c}$ cannot be obtained from direct evaluation of Eq. (20). In particular, making use of algebraic and graphical techniques, Bondi (1952) obtained values of $\lambda_c = 1.12$ for $\gamma = 1$ and $\lambda_c = 0.25$ for $\gamma = 5/3$. A useful fitting formula for $\lambda_c$ as a function of $\gamma$ was obtained by Fukue (2001) as

$$\lambda_c = -\frac{5}{4}\gamma + \frac{19}{8}, \tag{23}$$

which gives reasonably accurate values of $\lambda_c$ in the interval $1 \le \gamma \le 5/3$ compared to Eq. (20). The Bondi solution is then obtained by solving Eqs. (17) and (18) in coupled form for the density $\rho_B(r)$ and velocity $v_B(r)$ as a function of the radial distance from the source ($r = 0$). These profiles will then be used as initial conditions for the numerical simulations.

### 3.3 Model set-up

The simulations start from a spherical shell of gas centred on the SMBH as shown schematically in Fig. 1. The shell has inner radius $r_{in} = 0.02$ pc and outer radius $r_{out} = 10$ pc. The mass of the SMBH is set to $M_{BH} = 10^8\ M_\odot$, while the mass of the gas shell is $10^7\ M_\odot$. At very small radii the general-relativistic gravitational field effects are approximated by representing the SMBH by the pseudo-Newtonian Paczyńsky & Wiita (1980) potential given by

$$\phi_{PW} = -\frac{GM_{BH}}{r - R_g} \quad \text{with} \quad R_g = \frac{2GM_{BH}}{c^2}, \tag{24}$$

where $R_g$ is the gravitational radius of the black hole (Barai et al. 2011). The influence of this gravitational field on the accreting gas is represented by means of a static potential approach, where the radial acceleration

$$\mathbf{a}_{PW} = -\frac{GM_{BH}}{(r - R_g)^2}\hat{\mathbf{r}}, \tag{25}$$

is added to each particle.

The spherical computational domain $r_{in} \le r \le r_{out}$ is initially filled with particles distributed in a glass-like configuration (Couchman et al. 1995). In order to fit the analytical radial density Bondi profile the particles are then stretched so that they become more concentrated towards the centre. The initial density and radial velocity profiles are taken from the Bondi solution such that $\rho_i(r) = \rho_B(r)$ and $\mathbf{v}_i(r) = (v_{r,i}(r) = v_B(r), v_{\theta,i} = 0, v_{\phi,i} = 0)$, where standard transformations from spherical to Cartesian coordinates are operated to convert the velocity components into Cartesian form $\mathbf{v} = (v_x, v_y, v_z)$. As in Barai et al. (2011), the initial conditions are generated using an adiabatic index $\gamma_i = 1.01$ in Eq. (18), while the simulations are run with $\gamma_r = 1.0$ for the isothermal case and $\gamma_r = 5/3$ for the adiabatic case.

### 3.4 Boundary conditions

Particles crossing the inner radius $r_{in}$ are absorbed and counted in the mass inflow rate. Therefore, the region inside $r_{in}$ works as a static sink. The absorbed particles are removed from the sink and placed at some external radius given by $r_{out} - \epsilon\langle\Delta r\rangle$ with zero radial velocity and pressure gradient, where $\epsilon$ is a random number in the interval $(0, 1)$ and $\langle\Delta r\rangle$ is the mean particle separation calculated over the whole domain. This preserves the angular coordinates of the absorbed particles and enforces spherical symmetry and mass conservation. However, a future suite of calculations will consider the effects of adding the accreting mass to the central sink, thus allowing the mass of the SMBH to vary with time. In addition, inflow particles on or near $r_{in}$ have kernel supports that overlap with the sink and so they will miss some neighbours, causing the density to deviate from the Bondi profile. This deficiency can be in principle corrected by simply replacing the standard summation interpolant for the density

$$\rho_a = \sum_{b=1}^{n_{neigh}} m_b W_{ab}, \tag{26}$$

by a zeroth-order, Shepard interpolant of the form

$$\rho_a = \frac{\sum_{b=1}^{n_{neigh}} m_b W_{ab}}{\sum_{b=1}^{n_{neigh}} \frac{m_b}{\rho_b} W_{ab}}, \tag{27}$$

whenever a particle $a$ happens to be at a distance $< h$ from $r_{in}$. However, for the calculations with the consistent GADGET-2 code $h$ at the centre is always greater than $r_{in}$, even at the highest resolution, and so the effects of particle deficiency near the inner boundary are negligible.

In order to correctly reproduce the Bondi accretion a treatment of the outer boundary is also required. As shown by Barai et al. (2011), particles close to $r_{out}$ will feel an outward pressure gradient because of missing neighbours outside $r_{out}$. This finite pressure gradient makes the particles to revert the sign of their velocities in the proximity of $r_{out}$ and to flow out the computational domain. At higher temperatures, the pressure gradient will be stronger and eventually the net mass flux will be dominated by the outflow of particles across $r_{out}$. As a consequence, the continuous decrease of particles between $r_{in}$ and $r_{out}$ leads to an insufficient amount of mass to correctly describe the Bondi accretion rate. To remedy this SPH outflow problem, we implement one of the strategies proposed by Barai et al. (2011) (their run 7b). That is, the pressure gradient in the momentum equation is artificially set to zero for all particles lying within a thin shell of width $0.1r_{out}$ adjacent to the outer radius and for those particles having a positive radial velocity ($v > 0$) in an outer shell of width $0.6r_{out}$. With this provision, the outflowing mass is drastically reduced and only about 1% of the particles escapes from the simulation volume through $r_{out}$. However, exploratory tests with the isothermal models have shown that this small number of escaping particles has an adverse effect on the mass accretion rate at $r_{in}$, causing it to reproduce a steady-state accretion for some time before falling off at later times. In order to improve on this feature an additional more stringent condition is applied at the outer boundary by setting to zero the velocity of all particles still having $v > 0$ within the outermost shell of width $0.1r_{out}$. With this realization no particles leave the simulation volume at $r_{out}$ and steady-state Bondi accretion is prolonged in time.

### 3.5 The GIZMO code

Due to the lack of a true solution of Bondi accretion in 3D and in order to assess the performance of our modified SPH method, we compare the results with those from identical models using a public version of the GIZMO code (Hopkins 2015). This code is known to be second-order accurate (i.e., $C^1$-consistent) and has been designed to capture advantages of both the SPH method and the grid-based/adaptive mesh refinement (AMR) scheme of Berger & Colella (1989). In GIZMO the SPH inconsistency is corrected by normalizing the kernel such that the sum of fractional weights is always unity at every point and therefore there is no need to accommodate large numbers of neighbours within the compact support of the kernel. A complete account of the method is given by Hopkins (2015) and here we shall only mention the prescriptions that have been adopted to construct identical accretion models for comparison with our consistent version of the GADGET-2 code.

For the comparative calculations of this paper we use the PSPH mode of GIZMO, which employs a pressure-based formulation of the equations of motion. We follow the standard practice of using $n_{\rm neigh} = 64$ with a cubic $B$-spline kernel and a Wendland $C^4$ function. Unlike our modified SPH formulation, the artificial viscosity is evaluated using the method of Cullen & Dehnen (2010) with the variation of the artificial viscosity coefficient as given by the default setting $0.05 \leq \alpha_0(t) \leq 1.1$. In order to maintain consistency with our GADGET-2 models, the artificial conductivity and the particle splitting options were disabled, while the pseudo-Newtonian Paczyńsky & Wiita (1980) potential and the particle excision options were activated. As a further comparison test, we employ the meshless finite-mass (MFM) implementation of GIZMO, which is formally a Lagrangian-Eulerian finite-volume Godunov scheme. In this method, the particles are just moving cells, representing finite volumes with a well-defined volume partition, while gradient estimators are free from the kernel gradients (e.g., see Hopkins 2015). The only modification to the GIZMO code consists of a separate routine that has been added to implement the same outer boundary conditions as described in Section 3.4. Since in GIZMO the pressure is defined by the relation $p = (\gamma - 1)U\rho$, where $U$ is the specific internal energy, we must set $\gamma_{\rm r} = 1.001$ for the isothermal runs in order to avoid a zero pressure.

### 3.6 Model parameters

The list of runs and parameters are shown in Table 1. We consider four separate sets of calculations with increasing number of total particles ($N$). The set of models B1C$_{\rm ISO}$–B6C$_{\rm ISO}$ ($\gamma = 1$) and B1C$_{\rm ADIA}$–B6C$_{\rm ADIA}$ ($\gamma = 5/3$) were run with the standard GADGET-2 code with a fixed number of neighbours ($n_{\rm neigh} = 64$) independently of $N$, while models B1W$_{\rm ISO}$–B6W$_{\rm ISO}$ and B1W$_{\rm ADIA}$–B6W$_{\rm ADIA}$ were calculated with our modified code. Out of the family of possible curves describing the dependence of $n_{\rm neigh}$ on $N$, we choose the scaling relations $n_{\rm neigh} \approx 7.61N^{0.5}$ and $h \approx 15.23n_{\rm neigh}^{-0.34}$. These scalings are derived by requiring that $h/n_{\rm neigh} = 2N^{-2/3}$, which accommodates large numbers of neighbours for given $N$ while keeping reasonably large values of $h$. According to the parameterization $n_{\rm neigh} \sim N^{1-3/\beta}$, an exponent of $\approx 0.5$ corresponds to $\beta \approx 6$, which is an intermediate value in the interval $5 \leq \beta \leq 7$ (valid for low-order kernels as the cubic $B$-spline and the Wendland $C^4$ function for which the kernel approximation has a truncation error $\propto h^2$) and is appropriate for regular and quasi-ordered particle distributions for which the particle approximation error goes as $n_{\rm neigh}^{-1}$. With this choice, $h$ varies with $N$ as $h \approx 7.64N^{-0.17}$ so that as $n_{\rm neigh}$ increases with resolution, the smoothing length decreases, asymptotically approaching the limit $h \to 0$ when $n_{\rm neigh} \to \infty$ and $N \to \infty$ as required to restore particle consistency.

In order to test the convergence rate with varying $n_{\rm neigh}$, we add an extra simple routine in the code to evaluate the quality of the discrete moments $M_{0,a}$ and $\mathbf{M}'_{1,a}$ defined by the sums in Eqs. (13) and (14), respectively. Coupled to these we also evaluate the quality of the additional moments

$$\mathbf{M}_{1,a} = \sum_{b=1}^{n_{\rm neigh}} \frac{m_b}{\rho_b} \mathbf{r}_{ba} W_{ab}, \tag{28}$$

$$\mathbf{M}'_{0,a} = \sum_{b=1}^{n_{\rm neigh}} \frac{m_b}{\rho_b} \nabla_a W_{ab}, \tag{29}$$

and

$$\mathbf{M}'_{2,a} = \sum_{b=1}^{n_{\rm neigh}} \frac{m_b}{\rho_b} \mathbf{r}_{ba}^2 \nabla_a W_{ab}, \tag{30}$$

which should vanish because of the symmetry of the kernel function. The degree of consistency and its dependence on $n_{\rm neigh}$ is measured by the quality of the discrete moments, i.e., by tracking how well the kernel consistency relations are reproduced by the particle approximation.

Also listed in Table 1 are the parameters used in the GIZMO calculations for the isothermal Bondi accretion with two different resolutions and $n_{\rm neigh} = 64$. Models (B4GIZ$_{\rm (CS-ISO)}$, B4GIZ$_{\rm (WD-ISO)}$) and (B6GIZ$_{\rm (CS-ISO)}$, B6GIZ$_{\rm (WD-ISO)}$) were run with the PSPH mode of GIZMO with $N = 126^3$ and $200^3$ particles, respectively, while models B4GIZ$_{\rm (FM-ISO)}$ and B6GIZ$_{\rm (FM-ISO)}$ were run using the MFM formulation of GIZMO with $N = 126^3$ and $200^3$ particles, respectively.

## 4 RESULTS

In this section we describe the results obtained with our consistent version of the GADGET-2 code for the 3D spherical Bondi accretion onto a SMBH. These results are compared with identical simulations using the standard GADGET-2 and the GIZMO codes with fixed $n_{\rm neigh}(= 64)$.

### 4.1 Isothermal models

For the isothermal models we take $\rho_\infty = 10^{-19}$ g cm$^{-3}$ and $T_\infty = 10^7$ K. Figures 2, 3, and 4 show density maps and velocity vectors in the midplane for the more representative models B6C$_{\rm ISO}$, B6W$_{\rm ISO}$, and B6GIZ$_{\rm (FM-ISO)}$ (see Table 1), respectively, at 14.8 kyr during the gas accretion process. Only the inner region is shown in each plot. At this time ($\approx 1.69t_{\rm B}$), a steady-state accretion has already been established in all cases. We see large differences in the density distribution close to $r_{\rm in} = 0.02$ pc between model B6W$_{\rm ISO}$ and models B6C$_{\rm ISO}$ and B6GIZ$_{\rm (FM-ISO)}$. In particular, the cross-section images show that spherical symmetry is lost for models B6C$_{\rm ISO}$ and B6GIZ$_{\rm (FM-ISO)}$ in the densest shells close to $r_{\rm in}$ (see discussion in Section 5). In contrast, spherical symmetry is well maintained in the inner shells of model B6W$_{\rm ISO}$, which exhibits lower central densities compared to the other two models. This occurs because at comparable $N$ the sequence of models with fixed $n_{\rm neigh}$ has associated values of $h$ that are smaller than models working with large $n_{\rm neigh}$. In this case, the velocity vectors are pointing inwards symmetrically with magnitudes that are underestimated compared to the exact Bondi solution for $r \lesssim 0.1$ pc (see Fig. 5 and discussion below). However, a direct quantitative comparison between these two sets of models is not possible because working with fixed $n_{\rm neigh} = 64$ as $N$ increases results in much smaller values of $h$ than those obtained by scaling $n_{\rm neigh}$ with the same values of $N$. Thus, working consistently with comparable values of $h$ would require recalculating model B6W$_{\rm ISO}$ with much larger $n_{\rm neigh}$, which, on the other hand, would demand using also much larger values of $N$ compared to models B6C$_{\rm ISO}$ and B6GIZ$_{\rm (FM-ISO)}$. However, finer values of $h$, while losing consistency, does not imply higher accuracy for the particle approximation (see Eq. 15). Therefore, a comparison between both sets of models must be done in terms of the quality of the consistency relations (13), (14), and (28)-(30).

Figure 5 shows the radial variations of the density, radial component of the velocity, and Mach number within the interval $r_{\rm in} \leq r \leq r_{\rm out}$ for models B1W$_{\rm ISO}$–B6W$_{\rm ISO}$. For comparison the profiles of the more representative GIZMO models B6GIZ$_{\rm (CS-ISO)}$, B6GIZ$_{\rm (WD-ISO)}$, and B6GIZ$_{\rm (FM-ISO)}$ are also shown. The solid black line in the left and right top panels corresponds to the exact Bondi profiles. The symbols describe the numerical solutions and correspond to the median (i.e., the maximum) of the distributions. In all cases, with the exception of models B1W$_{\rm ISO}$–B2W$_{\rm ISO}$, the bounds of the 90th percentile of the distributions are so close to the median that they are not shown in Fig. 5 for the sake of clarity. The density and radial velocity profiles of models B1W$_{\rm ISO}$–B6W$_{\rm ISO}$ closely match the exact Bondi solution, except near the inner and outer radii. At $r \lesssim 0.1$ pc, the density is underestimated in model B6W$_{\rm ISO}$. The deviations from the exact density profile become progressively larger and extend in radius up to $\approx 0.2$ pc as the resolution is decreased (models B1W$_{\rm ISO}$–B5W$_{\rm ISO}$). The radial velocity profiles match the Bondi solution for radii $r \gtrsim 0.8$ pc (for model B1W$_{\rm ISO}$) and $r \gtrsim 0.3$ pc (for model B6W$_{\rm ISO}$) within which the inflowing velocities are slightly underestimated compared to the exact solution. As was pointed out by Barai et al. (2011), the deviations of the numerical profiles from the exact Bondi solution close to the centre are an artefact of the inner boundary conditions due to truncation of the kernel for particles close to $r_{\rm in}$. However, in the case of models B1W$_{\rm ISO}$–B6W$_{\rm ISO}$ the deviations near the inner radius are due to oversmoothing because of their larger associated values of $h$. For instance, for model B6W$_{\rm ISO}$ the values of $h$ close to the centre are $\approx 10 r_{\rm in}$ compared to B6C$_{\rm ISO}$ and the GIZMO models where $h$ is $\approx r_{\rm in}$ and $\approx 0.6 r_{\rm in}$, respectively. The oversmoothing in model B6W$_{\rm ISO}$ clearly suppresses most of the non-axisymmetric noise present in models B6C$_{\rm ISO}$ and B6GIZ$_{\rm (FM-ISO)}$ around the accretor (see Figs. 2 and 4). As the spatial resolution is increased, the number of neighbours within the kernel support increases and the smoothing length decreases. As a result, particle consistency is restored and therefore the spatial extent and magnitude of the deviation decrease as we may see by comparing models B1W$_{\rm ISO}$ and B6W$_{\rm ISO}$. In terms of the percent root-mean-square error (RMSE), the deviations of the density and velocity profiles from the Bondi solution in the range $0.02 \leq r \leq 9$ pc go from ($\approx 85\%$, $\approx 21\%$) for model B1W$_{\rm ISO}$ to ($\approx 10\%$, $\approx 9\%$) for model B6W$_{\rm ISO}$. Thus, increasing $N$ and $n_{\rm neigh}$ lessens the magnitude of the errors, consistently with Eq. (15).

On the other hand, the rapid decrease of the density at $r \gtrsim 9$ pc is associated to the particle deficiency at the outer boundary. As stated by Nagamine et al. (2012), this problem is related to difficulties of setting appropriate outer boundary conditions in SPH since it does not arise in grid methods where the outer boundary is fixed by specifying $\rho_\infty$ and $T_\infty$. To remedy the outflow of particles at $r_{\rm out}$, the pressure gradient for all those particles having a positive radial velocity in the range $6 \leq r \leq 9$ pc is artificially zeroed, while the velocity of those particles still having a positive radial velocity within $9 < r \leq 10$ pc is also zeroed. This latter condition prevents particles to flow out the simulation volume, while the former condition causes the velocity field to slightly oscillate at radii $r \gtrsim 4$ pc. However, the amplitude of the oscillations is seen to decrease as the spatial resolution is increased because less particles are reverting its velocity from negative to positive as the effects of the kernel deficiency at $r_{\rm out}$ become less important.

Finally, Fig. 5c depicts the Mach number, $\mathcal{M} = |v|/c_{\rm s}$, as a function of radius. For all models, the location of the sonic point is $R_{\rm s} \approx 1.67$ pc regardless of the spatial resolution. This is consistent with Eq. (21) for the Bondi solution, which predicts $R_{\rm s} \approx 1.6$ pc. Also independently of the spatial resolution models B1C$_{\rm ISO}$–B6C$_{\rm ISO}$ predicted $R_{\rm s} \approx 1.7$ pc.

For comparison models B6GIZ$_{\rm (CS-ISO)}$, B6GIZ$_{\rm (WD-ISO)}$, and B6GIZ$_{\rm (FM-ISO)}$ exhibit similar density and radial velocity profiles, following the trends of the Bondi solution. However, in all three cases the density and radial velocity are both overestimated at almost all radii. Similar plots were also obtained for models B4GIZ$_{\rm (CS-ISO)}$, B4GIZ$_{\rm (WD-ISO)}$, and B4GIZ$_{\rm (FM-ISO)}$. In terms of the RMSE, the deviations of the density and velocity profiles across the interval $0.02 \leq r \leq 9$ pc amount to ($\approx 358\%$,$\approx 23\%$) for B6GIZ$_{\rm (CS-ISO)}$, ($\approx 609\%$,$\approx 23\%$) for B6GIZ$_{\rm (WD-ISO)}$, and ($\approx 821\%$,$\approx 24\%$) for B6GIZ$_{\rm (FM-ISO)}$, which are significantly larger compared with model B6W$_{\rm ISO}$. The density and velocity radial profiles obtained with GIZMO are similar to those for model B6C$_{\rm ISO}$, implying that normalization of the kernel to correct for particle deficiencies close to $r_{\rm in}$ is not enough to restore complete consistency since it may be equally compromised by particle disorder and varying smoothing lengths. This could explain the difference with the radial profiles of model B6W$_{\rm ISO}$. Therefore, it would be interesting to investigate the response of GIZMO under large numbers of neighbours, which will demand implementing some modifications to the original code. Moreover, it can be seen from Fig. 5c that for all GIZMO models the sonic point is predicted at $\approx 3.2$ pc, which is a factor of two larger than the Bondi value of 1.6 pc. In contrast to model B6C$_{\rm ISO}$, spherical symmetry is better preserved by the GIZMO models in the densest shells close to $r_{\rm in}$ (see Section 5). The velocity vectors are seen to point inwards with magnitudes that are similar to those shown in Fig. 2 for B6C$_{\rm ISO}$. Towards the outer boundary the radial velocity is well-bahaved, except for small oscillations at radii $\gtrsim 8$ pc, whereas the density decreases sharply at $r \approx 7$ pc and then increases rapidly beyond $r = 9$ pc, reaching relatively high values at $r_{\rm out} = 10$ pc. This is an effect of the outer boundary conditions implemented that do not allow matter to flow out the computational volume, causing some mass to accumulate close to $r_{\rm out}$. This effect appears to be much more pronounced in the GIZMO models than in the standard and modified GADGET-2 models.

Figures 6 and 7 show the first two moments of the kernel function, as defined by Eqs. (13) and (28), and the first moments of the gradient of the kernel, as defined by Eqs. (14), (29), and (30), for models B1W$_{\rm ISO}$–B6W$_{\rm ISO}$ and B1C$_{\rm ISO}$–B6C$_{\rm ISO}$, respectively. We may see from Fig. 6 that as $n_{\rm neigh}$ is increased, the distributions of $M_{0,a}$ and the mean of the components of $\mathbf{M}'_{1,a}$, namely $\langle \mathbf{M}'_{1,a} \rangle$, both approach a Dirac-$\delta$ distribution with peaks becoming closer and closer to $\approx 1$, while the distributions of the other moments peak at zero with very good accuracy, meaning that the symmetry of the kernel is very well reproduced by the particle approximation. Thus, $C^0$ and $C^1$ particle consistencies are restored in the sequence of models B1W$_{\rm ISO}$–B6W$_{\rm ISO}$ as $n_{\rm neigh}$ is increased, implying approximate second-order accuracy for the simulations. As shown in Fig. 7, the distributions of $M_{0,a}$ and $\langle \mathbf{M}'_{1,a} \rangle$ for models B1C$_{\rm ISO}$–B6C$_{\rm ISO}$ are significantly much broader, approaching a Gaussian shape with peaks at $\approx 0.83$ and $\approx 0.996$, respectively, almost independently of $N$. Although, the distributions of the other moments have also peaks pointing very close to zero, the deviations of $M_{0,a}$ and $\langle \mathbf{M}'_{1,a} \rangle$ from unity imply a loss of $C^0$-consistency for these models, meaning that the simulations are not even first-order accurate.

### 4.2 Mass inflow rates for the isothermal models

The mass inflow rate, or accretion rate, is calculated by summing up the masses of all the accreting particles crossing the inner boundary, $r_{\rm in} = 0.02$ pc, during the time interval $\Delta t$ so that

$$\dot{M}_{\rm acc}(t) = \frac{1}{\Delta t} \sum_{(r_k < r_{in})} m_k, \tag{31}$$

where the sum is over all particles of mass $m_k$ at radii $r_k < r_{in}$ that have fallen inside the inner boundary. Here we set $\Delta t = 0.17$ kyr. Figure 8 shows the mass inflow rates as a function of time for models B1C$_{\rm ISO}$–B6C$_{\rm ISO}$ (top panel) and models B1W$_{\rm ISO}$–B6W$_{\rm ISO}$ (bottom panel). Also shown in the bottom panel are the resulting mass inflow rates from the highly resolved GIZMO models. The horizontal dashed line in each panel marks the Bondi accretion rate, $\dot{M}_{\rm B} \approx 80.8\ M_\odot$ yr$^{-1}$ ($\approx 5.1 \times 10^{27}$ g s$^{-1}$), as calculated from Eq. (19) with $\gamma = 1.01$. After a sharp increase at the beginning, steady-state accretion is quickly achieved by models B1C$_{\rm ISO}$–B6C$_{\rm ISO}$ with temporal average values of $\approx 20.1\ M_\odot$ yr$^{-1}$ for model B1C$_{\rm ISO}$ and $\approx 8.9\ M_\odot$ yr$^{-1}$ for model B6C$_{\rm ISO}$ (see Table 1). These are factors from 4 to 8 times lower than the exact Bondi rate. In spite of this discrepancy, an almost flat profile for $\dot{M}_{\rm acc}(t)$ is very well reproduced by these simulations, with convergence to an almost constant $\approx 10\ M_\odot$ yr$^{-1}$ rate being achieved by models B3C$_{\rm ISO}$–B6C$_{\rm ISO}$. These results contrast with those obtained by Barai et al. (2011) for similar simulations with the Bondi initial conditions and $r_{\rm out} = 10$ pc but larger $r_{\rm in}$(= 0.1pc), where the Bondi accretion was reproduced for some time duration before decaying at longer times.

With the consistent SPH code, models B1W$_{\rm ISO}$ and B2W$_{\rm ISO}$ with lower resolution were also reproducing approximately constant rates during the first 15 kyr, with average values which were factors of $\sim 2$ higher than the Bondi rate. As the resolution is increased (models B3W$_{\rm ISO}$–B6W$_{\rm ISO}$), the results converge to an approximate Bondi plateau for the first 14 kyr before suddenly rising to values as high as $\approx 100\ M_\odot$ yr$^{-1}$. The time average values of the mass inflow rates at $r_{\rm in} = 0.02$ pc for these models are also listed in Table 1. We may see that in terms of a percent relative error the time-averaged numerical accretion rates deviate from the Bondi rate from $\approx 0.97\%$ (for model B3W$_{\rm ISO}$) to $\approx 0.21\%$ (for model B6W$_{\rm ISO}$). Evidently, restoring $C^1$-consistency for the particle approximation translates into accretion rates that closely match the analytical solution. This is in strong contrast with models B3C$_{\rm ISO}$–B6C$_{\rm ISO}$, which are not even $C^0$-consistent. On the other hand, the GIZMO models are all below the Bondi rate with average deviations of $\approx 49\ M_\odot$ yr$^{-1}$ for B6GIZ$_{\rm (CS-ISO)}$, $\approx 60\ M_\odot$ yr$^{-1}$ for B6GIZ$_{\rm (WD-ISO)}$, and $\approx 46\ M_\odot$ yr$^{-1}$ for B6GIZ$_{\rm (FM-ISO)}$. In all three cases the mass inflow rate increases slowly with time, eventually reaching a steady-state accretion at late times ($\gtrsim 15$ kyr). No decay of the accretion rate was observed at longer times in any of the models. This is consistent with the fact that there are no particles leaving the simulation volume through $r_{\rm out}$. This way enough particles are maintained between $r_{\rm in}$ and $r_{\rm out}$ to correctly describe a fairly constant mass accretion rate. The scatter in the data present in models B3C$_{\rm ISO}$–B6C$_{\rm ISO}$ and the low resolution cases B1W$_{\rm ISO}$ and B2W$_{\rm ISO}$ is strongly damped in models B3W$_{\rm ISO}$–B6W$_{\rm ISO}$ working with larger numbers of neighbours. In contrast, no scatter in the data is observed in any of the GIZMO models.

### 4.3 Adiabatic models

We now drop the isothermal assumption and examine the case where the temperature is allowed to vary adiabatically. As for the isothermal models we use $\rho_\infty = 10^{-19}$ g cm$^{-3}$, $T_\infty = 10^7$ K, and $M_{\rm BH} = 10^8\ M_\odot$ with $\gamma_{\rm r} = 5/3$. We consider two separate sets of simulations with increasing resolution, all starting from initial conditions corresponding to the isothermal Bondi solution. Models labelled B1C$_{\rm ADIA}$–B6C$_{\rm ADIA}$ were run with the standard GADGET-2 code with fixed $n_{\rm neigh}(= 64)$, while models B1W$_{\rm ADIA}$–B6W$_{\rm ADIA}$ were calculated with the modified code (see Table 1).

Figures 9 and 10 show midplane density (top) and temperature (bottom) images and velocity vectors of the inner regions for the more representative models B6C$_{\rm ADIA}$ and B6W$_{\rm ADIA}$ at 14.8 kyr, respectively. A hot ($k_{\rm B}T \sim 2$–8 keV) and dense spherical shell is evident around the SMBH. In both cases, the densest inner shells appear to be thicker than in their isothermal counterparts of Figs. 2 and 3. The absence of perceptible scatter in the velocity vectors in Fig. 10 shows that the inflow is kept nearly spherically symmetric for model B6W$_{\rm ADIA}$ compared to model B6C$_{\rm ADIA}$ where spherical symmetry is lost due to the presence of small density and temperature fluctuations in the hottest and densest shells near $r_{\rm in}$. Emission of hard X-ray photons as observed in real systems with the new generation of X-ray telescopes (Braito et al. 2007; Farrell et al. 2009; Dewangan et al. 2013; Russell et al. 2015; Luangtip et al. 2016) may well be originating in these hot and dense regions.

The variations of the density, radial component of the velocity, Mach number, and temperature with radius at 15 kyr are shown in Fig. 11 for models B1W$_{\rm ADIA}$–B6W$_{\rm ADIA}$. The density, velocity, and Mach number profiles are quali-tatively similar to those for the isothermal Bondi accretion of Fig. 5. For reference the exact isothermal Bondi profiles (solid black lines) are depicted for the density and radial velocity (top panels). The solid red lines depict the limiting expressions for the density, radial velocity, and temperature for $r \ll R_B$ in a highly collisional Bondi accretion for $\gamma = 5/3$ (Sari & Goldreich 2006), i.e.,

$$\rho \sim \frac{\rho_\infty}{8}\left(\frac{2R_B}{r}\right)^{3/2}, \tag{32}$$

$$v \sim -\frac{c_{s,\infty}}{2}\left(\frac{2R_B}{r}\right)^{1/2}, \tag{33}$$

$$T \sim \frac{T_\infty}{4}\left(\frac{2R_B}{r}\right). \tag{34}$$

Since the adiabatic runs start from initial conditions corresponding to the isothermal Bondi solution, the density and radial velocity follow approximately the isothermal Bondi profile at large radii ($r \gtrsim 0.4$ pc for the density and $r \gtrsim 2$ pc for the velocity), while the adiabatic Bondi solution near the inner boundary (solid red lines) is shifted towards lower densities and smaller velocities compared to the isothermal solution due to adiabatic heating close to $r_{in}$, which slows down the radial infall of matter into the SMBH. The numerical density profile for model B6W$_{ADIA}$ deviates from the isothermal Bondi solution near $r_{in}$. At decreased spatial resolution the deviations increase in magnitude and radial extension. This feature is a combined effect of the larger smoothing lengths associated with the lower resolution runs for which oversmoothing at $r_{in}$ is greater and the rapid raise of the temperature inside $\approx 0.5$ pc, as shown in Fig. 11d. On the other hand, as expected, the inflow radial velocities are smaller than for the isothermal accretion as shown by the large deviations from the Bondi profile at radii $r \lesssim 2$ pc. However, by extrapolating the solid red curve for the temperature to large radii it would appear that the numerical profiles are fairly matching the adiabatic Bondi solution at least up to $\approx 0.1$ pc, when significant departures are evident towards smaller radii. In addition, smaller velocities are achieved in models B5W$_{ADIA}$ and B6W$_{ADIA}$ compared to the lower resolution runs. This occurs essentially because in the latter models the temperature inside $r \approx 0.5$ pc rises faster than in the lowest resolution models. For instance, at $\approx 0.1$ pc the temperature is $\approx 1.3 \times 10^8$ K for model B1W$_{ADIA}$ against $\approx 2.0 \times 10^8$ K for model B6W$_{ADIA}$. This produces pressure gradients that are correspondingly larger in the high resolution models, thereby leading to smaller velocities. In passing, we note that the rate of temperature increase for models B5W$_{ADIA}$ and B6W$_{ADIA}$ near the inner boundary closely follow that predicted by the adiabatic Bondi solution. However, the numerical solutions predict higher temperatures than the Bondi solution at comparable radius. This shift in temperature is related to the shift in density between the isothermal and adiabatic Bondi solution since for $\gamma = 5/3$ the temperature varies as $\sim \rho^{2/3}$.

The adiabatic sound speed for $\gamma_r = 5/3$ is about 1.29 times the value for the isothermal models. This causes the Bondi radius $R_B$ to decrease from $\approx 3.25$ pc for the isothermal case to $\approx 1.97$ pc when $\gamma = 5/3$, while the sonic radius $R_s = 0$ according to relation (21). From the Mach number profiles we may see that the sonic point is at $r \approx 0.3$ pc with the exception of model B1W$_{ADIA}$ for which the transition to supersonic flow occurs at $r \approx 0.5$ pc. Inside this radius and deep into the inner zones the gas flows supersonically however at a slower rate compared to the isothermal models. The temperature profiles displayed in Fig. 11 are qualitatively similar for all models, except for $r \lesssim 0.2$ pc where the temperature rises slightly faster for the higher resolution models, reaching values of $\approx 3.0 \times 10^8$ K close to $r_{in} = 0.02$ pc. At radii $r \gtrsim 0.5$ pc, the temperature decays ($k_B T \approx 0.2$–1 keV) from $T \sim 10^8$ K close to the sink border to $T \approx 10^{6.5}$ K at $r_{out}$. This is typical of the so-called warm absorber features seen in the soft X-ray spectra of AGNs (Dewangan et al. 2007; Ramírez 2008, 2013). The time variation of the mass accretion rates at $r_{in} = 0.02$ pc is shown in Fig. 12 for models B1C$_{ADIA}$–B6C$_{ADIA}$ (top panel) and models B1W$_{ADIA}$–B6W$_{ADIA}$ (bottom panel). In this case, we have set $\Delta t = 0.51$ kyr in Eq. (31). The horizontal dashed line in each panel marks the exact Bondi rate $\dot{M}_B \approx 8.6\ M_\odot$ yr$^{-1}$. In both sets of models approximate steady state accretion is reached after about 10 kyr, when the profiles become progressively flatter. Considerable scatter is present in the data for the lower resolution models in both cases. As the resolution is increased the scatter is significantly reduced and the accretion rate approaches a much smoother behaviour. The time average accretion rates vary from $\approx 4.4\ M_\odot$ yr$^{-1}$ for model B1C$_{ADIA}$ to $\approx 1.4\ M_\odot$ yr$^{-1}$ for model B6C$_{ADIA}$ (see Table 1), which are factors of $\approx 2$ to $\approx 6$ lower than the Bondi accretion. In contrast, models B1W$_{ADIA}$–B6W$_{ADIA}$ reproduce accretion rates that closely match the Bondi accretion (after $\approx 10$ kyr) for the higher resolution models (B5W$_{ADIA}$ and B6W$_{ADIA}$). For these latter models the time averaged accretion rates deviate from the Bondi rate by $\approx 12\%$ (model B5W$_{ADIA}$) to $\approx 9\%$ (model B6W$_{ADIA}$).

Finally, Figs. 13 and 14 display the distributions of the moments of the kernel and its gradient for models B1W$_{ADIA}$–B6W$_{ADIA}$ and B1C$_{ADIA}$–B6C$_{ADIA}$, respectively, after 15 kyr. In Fig. 13, the distributions of $M_{0,a}$ and $\langle \mathbf{M}'_{1,a} \rangle$ are seen to approach a Dirac-$\delta$ distribution with peaks very close to $\approx 1$ for model B6W$_{ADIA}$ working with $n_{neigh} = 22512$. As the number of neighbours increases with the spatial resolution, the moments $M_{0,a}$ and $\langle \mathbf{M}'_{1,a} \rangle$ gets closer and closer to 1, implying $C^0$ particle consistency. The distributions of the other moments all peak around zero, meaning that the symmetry properties of the kernel are numerically reproduced with good accuracy, with these moments also approaching a Dirac-$\delta$ distribution for large values of $n_{neigh}$. Conversely, models B1C$_{ADIA}$–B6C$_{ADIA}$, working with fixed $n_{neigh} = 64$, display moment distributions that are much broader, approaching Gaussian shapes with peaks that deviate from the required consistency constraints (i.e., $M_{0,a} \approx 0.83$ and $\langle \mathbf{M}'_{1,a} \rangle \approx 0.995$). Evidently, a loss of $C^0$-consistency accompanies these latter models, which makes the simulations even worse than first-order accuracy. As long as $N$ and $n_{neigh}$ are increased along the sequence of models B1W$_{ADIA}$–B6W$_{ADIA}$, the particle approximation restores $C^1$-consistency and captures the same order of accuracy of the kernel approximation as the zeroth-order discretization errors vanish (Read et al. 2010).

## 5 INSTABILITY OF THE BONDI ACCRETION

Since the work of Cowie (1977), numerous simulations on Bondi accretion has been devoted to investigate the instability of the flow. A thorough review of published numerical simulations of Bondi flow in a totally absorbing accretor prior to 2005 is given by Foglizzo et al. (2005). Also the former overview of Bondi accretion by Edgar (2004) provides a discussion on the flow instability. These simulations can be divided into three main groups, namely plane accretion in 2D, 3D axisymmetric accretion, and full 3D accretion. Simulations of planar flow in 2D shows the more unstable behaviour, where the shock, which is attached to the accretor, moves from side to side in a flip-flop manner (Matsuda et al. 1987; Soker 1990, 1991; Livio et al. 1991; Matsuda et al. 1991; Shima et al. 1998; Pogorelov et al. 2000). Denoting by $r_a = 2GM/v_\infty^2$ the accretion radius of an accretor of mass $M$ and velocity $v_\infty$, most planar simulations reach values as low as $r_\star/r_a = 0.001$. For such small accretors the instability was found to be strongest for intermediate to high Mach numbers (Matsuda et al. 1987; Pogorelov et al. 2000). The flip-flop instability was more recently revisited by Blondin & Pope (2009) who found that for a relatively large accretor ($r_\star/r_a = 0.037$) planar accretion is unstable for $\gamma = 4/3$ and stable for $\gamma \geq 1.6$, while for $\gamma = 5/3$ the instability requires that $r_\star/r_a < 0.0025$. On the other hand, 3D axisymmetric flows were found to be stable (e.g., Shima et al. 1985; Sawada et al. 1989; Font & Ibáñez 1998; Pogorelov et al. 2000) with few exceptions (e.g., Fryxell et al. 1987), where the instability was found in the form of a vortex shedding in the von Kármán manner when the accretor was a non absorbing sphere. For an absorbing accretor, Koide et al. (1991) found a vortex shedding for $\mathcal{M} \geq 1.4$ and $r_\star/r_a = 0.005$.

The instability in full 3D spherical accretion seems to be present in almost all simulations (Foglizzo et al. 2005). The origin of the instability is a detached bow shock, ahead of the accretor, if $\mathcal{M} \geq 0.6$, the accretor is small enough ($r_\star/r_a = 0.01-0.02$), and $\gamma = 4/3$ and $5/3$ (Ruffert & Arnett 1994; Ruffert & Anzer 1995; Ruffert 1995, 1997). The instability has been also observed in quasi-isothermal ($\gamma \approx 1$) flow simulations (Ruffert 1996, 1999) and for accretors with $r_\star/r_a \geq 1$. However, the instability is never as strong as the flip-flop instability observed for planar flows. In more recent work, the instability has been described as an 'entropic-acoustic' instability, where entropy perturbations introduced by a shock propagate back to the shock via sound waves (Foglizzo & Tagger 2000; Foglizzo 2001). A linear analysis of spherical adiabatic Bondi accretion has showed that the flow is unstable against external non-radial perturbations (Kovalenko & Eremin 1998). A conclusion that can be drawn from all these simulations is that for relatively large sizes of the accreting object (i.e., $r_\star/r_a \geq 0.25$), the instability is either weak or does not develop.

Although the mechanisms that has been suggested to describe the various instabilities are physical, a number of numerical issues has been raised as responsible for triggering or damping the instabilities. For instance, besides the damping effect of numerical viscosity, the carbuncle instability (Robinet et al. 2000) in the region where the shock is parallel to the grid may favour the artificial generation of vorticity and entropy perturbations, which in turn can feed an advective-acoustic cycle. Conversely, insufficient grid resolution between the shock and the accretor may artificially damp vorticity and entropy waves (Foglizzo et al. 2005). Moreover, in numerical simulations the sonic surface ahead of the accretor is generally attached to the accretor itself. This explains why the size of the accretor plays an important role in the strength of the instability. Since the surface of the accretor is in contact with a subsonic flow, the advection of vorticity and entropy perturbations through the inner boundary condition may induce an artificial acoustic feedback from the accretor. A further aspect which was not considered by the authors of the existing simulations is that rotation, and hence non-axisymmetry, can be induced by pure radial motions through hydrodynamical coupling even for initially spherically symmetric radial infall (Boss 1989; Sigalotti 1994). In particular, close to the accretor where the radial velocity amplifies, hydrodynamical coupling may give rise to rotational velocities of finite amplitude which then may contribute to sizeable non-axisymmetry as the radial velocity increases.

Recent simulations of the Hoyle-Lyttleton accretion in 3D with improved spatial resolution over previous simulations by Blondin & Raymer (2012) shows that for a large accretor ($r_\star/r_a = 0.05$) the flow remains steady and the mass accretion rate relaxes to a constant value with variations no greater than 0.02%. When the size of the accretor is reduced to $r_\star/r_a = 0.01$, a quasi-periodic axisymmetric breathing mode appears and the mass accretion rate is modulated by the oscillation of the bow shock. In contrast to previous simulations, the simulations of Blondin & Raymer (2012) do not display rotational flow and remain almost entirely axisymmetric. One important conclusion from these simulations is that the intrinsic noise generated at the bow shock due to numerical effects is insufficient to significantly affect the stability of the flow and would likely have an effect similar to that of numerical noise. However, the presence of slight asymmetries near the accretor in the small accretor calculation follow the general trend observed in most 2D simulations, where a smaller accretor produces a less stable flow pattern. These results are in accordance with the suggestions derived from the simulations of Ruffert (1997, 1999) that large-scale density and velocity gradients across the upstream accretion column is necessary to induce a non-stationary flow.

In contrast to previous simulations, our SPH calculations deal with a stationary accretor and therefore no bow shock is formed. The simulations B1C$_{\rm ISO}$–B6C$_{\rm ISO}$ and B1C$_{\rm ADIA}$–B6C$_{\rm ADIA}$ with standard SPH and $n_{\rm neigh} = 64$ all show small-scale density and velocity gradients near the accretor which are the result of non-linear amplification of particle noise inherent in SPH, which leads to different flow patterns as the resolution is increased. This noise arises because mutually repulsive pressure forces between particle pairs do not cancel in all directions simultaneously, giving rise to non-radial velocities whose magnitudes are larger near the accretor. On the other hand, the use of standard artificial viscosity with a constant coefficient leads to spurious angular momentum advection in the presence of vorticity. Figure 15a displays the magnitudes of the numerically induced non-radial velocities as a function of radius for models B6C$_{\rm ISO}$ (blue triangles) and B6W$_{\rm ISO}$ (black open dots). The symbols correspond to average values of the non-radial velocities taken over concentric shells of width ≈ 0.033 pc. Spherical symme-try is very well maintained for model B6W$_{\rm ISO}$, while significant deviation from spherical symmetry in model B6C$_{\rm ISO}$ is evidenced by the large scatter in the data close to the accretor. Such deviations are due to non-linear amplification of numerical noise caused by the irreducible zeroth-order discretization errors when working with small numbers of neighbours. In spite of this, the flow is steady and the accretion rate remains remarkably constant for the isothermal case and tends to a constant value late in the evolution for the adiabatic accretion. As shown in Fig. 15b, the GIZMO simulations also show deviations from spherical symmetry, albeit at a smaller extent than model B6C$_{\rm ISO}$. However, in all cases, the intrinsic noise generated due to numerical effects is not sufficient to make the flow unstable in the proximity of the accretor. As particle consistency is enforced, the level of numerical noise should be drastically reduced as the zeroth-order errors decay.

## 6 CONCLUSIONS

In this paper we have examined the three-dimensional, spherically symmetric Bondi accretion onto a supermassive black hole (SMBH) of mass $10^8$ $M_\odot$ within a radial range of 0.02-10 pc, using a modified version of the smoothed particle hydrodynamics (SPH) GADGET-2 code (Springel 2005), which achieves approximate first-order consistency ($C^1$), i.e., second-order accuracy even in the presence of spatially and temporally varying smoothing lengths (Gabbasov et al. 2017). In contrast to previous SPH calculations of gas accretion onto a SMBH, the number of neighbours, $n_{\rm neigh}$, and the smoothing length, $h$, are initially allowed to depend on the total number of particles, $N$, according to the scaling relations $n \sim N^{1-3/\beta}$ and $h \sim N^{-1/\beta}$ with $\beta \approx 6$ (Zhu et al. 2015). These relations comply with the asymptotic limits $N \to \infty$, $h \to 0$, and $n_{\rm neigh} \to \infty$ with $n_{\rm neigh}/N \to 0$ for full SPH consistency (Rasio 2000). A Wendland C$^4$ function is used as the interpolation kernel to support large numbers of neighbours while suppressing numerical instabilities due to close pairing of particles. An improved switch for the artificial viscosity was also implemented which avoids excessive viscous dissipation in subsonically convergent flows and applies the same weight to the velocity divergence and vorticity (Hu et al. 2014).

The effects of improved SPH consistency is measured by comparing the results with the exact Bondi solution and the quality of the particle consistency constraints resulting from identical SPH models using the standard GADGET-2 code with fixed $n_{\rm neigh}(= 64)$ regardless of $N$. Two independent sequences of calculations were considered to study the ability of the method to reproduce the isothermal ($\gamma = 1$) and adiabatic ($\gamma = 5/3$) Bondi accretion with increased spatial resolution. The performance of our consistent SPH method was also compared with the PSPH and the meshless finite-mass (MFM) formulations of the GIZMO code for the more representative isothermal accretion models. The main results can be summarized as follows:

(1) With the modified GADGET-2 code the isothermal runs produce density and radial velocity profiles that closely match the exact Bondi solution, except near the inner boundary ($r_{\rm in} = 0.02$ pc), where the density and velocity are slightly underestimated due to oversmoothing there. As $C^1$-consistency is approached by increasing $n_{\rm neigh}$ and decreasing $h$ with increasing resolution, the magnitude and radial extension of the deviations from the Bondi profiles are seen to decrease. In terms of the root-mean-square error (RMSE), the profiles deviate from the Bondi solution by $\approx 85\%$ for the density and $\approx 21\%$ for the radial velocity when $N = 63^3$ and $n_{\rm neigh} = 3941$ to $\approx 10\%$ for both the density and the velocity in the highest resolution case with $N = 200^3$ and $n_{\rm neigh} = 22512$. For comparison, the highly resolved models with GIZMO follow the trends of the Bondi solution with RMSE deviations for the density of $\approx 358\%$ and $\approx 609\%$ for the PSPH runs working with a cubic $B$-spline kernel and a Wendland C$^4$ function with fixed $n_{\rm neigh} = 64$, respectively, while for the MFM formulation the density deviation is $\approx 821\%$. For all three GIZMO calculations the deviations from the Bondi velocity profile are $\approx 23-24\%$.

(2) The location of the sonic point predicted by the isothermal simulations is $R_{\rm s} \approx 1.67$ pc regardless of the spatial resolution, which is close to the theoretical value of $\approx 1.6$ pc for isothermal Bondi accretion. Values close to 1.7 pc were predicted by the set of models calculated with the standard GADGET-2 code. The GIZMO models predicted a sonic radius of $\approx 3.2$ pc independently of the formulation employed, which is a factor of two greater than the Bondi value.

(3) All isothermal models reach a steady-state accretion quickly which is maintained for more than 15 kyr in the evolution with both SPH methods, except for the lowest resolution models which exhibit a large scatter of the mass inflow rate. The prolonged steady-state accretion is related to our more stringent treatment of the outer boundary conditions. However, only the highest resolution runs reach approximate $C^1$-consistency and therefore they closely match the Bondi accretion with time averaged values that are within $\sim 1\%$ (relative error) of the Bondi value. The GIZMO models predict accretion rates that are always below the Bondi rate and that slowly grow with time, eventually reaching a steady-state after about 15 kyr.

(4) The variations of the density and velocity with radius for the adiabatic ($\gamma = 5/3$) models are qualitatively similar to the isothermal profiles, except that for adiabatic accretion the inflow velocities near the inner boundary are smaller compared to the isothermal runs. In general, as the resolution is increased the inflow velocities become progressively smaller. This occurs because in the highly resolved models the temperature inside $r \approx 0.5$ pc rises faster than in the models with lower resolution, leading to correspondingly larger pressure gradients and smaller radial velocities. Comparison with a limiting solution for the adiabatic ($\gamma = 5/3$) Bondi accretion shows that the rate of increase of the temperature near the inner boundary closely match that for the Bondi solution. However, at comparable radii the numerical simulations predict higher temperatures close to the source which is caused by a combination of the isothermal Bondi initial conditions assumed for the adiabatic runs, numerical resolution, and inner boundary conditions.

(5) In the adiabatic simulations, the sonic point is predicted at $r \approx 0.3$ pc, which represents a 30% deviation from the exact value of $R_{\rm s} = 0$, while the Mach number remains close to zero for $r \gtrsim 0.6$ pc. With the exception of the lowest resolution runs, this feature is almost independent of the spatial resolution and SPH method.

(6) Since the Bondi radius $R_{\rm B}$ decreases from $\approx 3.25$ pc for the isothermal case to 1.97 pc when $\gamma = 5/3$, the standard and consistent SPH methods agree in predicting, at least for the highest resolution runs, a smooth decrease of the mass accretion rate from the beginning as the temperature raises. Approximate steady-state accretion at the inner boundary is reached after $\approx 10$ kyr, when the temperature near the inner radius stabilizes. The lowest resolution runs with both methods exhibit a large scatter in the data and so the mass accretion rate varies erratically with time. All runs with the standard GADGET-2 code fail to match the value of the adiabatic Bondi accretion rate of $\approx 8.6\ M_{\odot}\ {\rm yr}^{-1}$, while again only the highest resolved models working with $N = 159^3$, $n_{\rm neigh} = 15927$ and $N = 200^3$, $n_{\rm neigh} = 22512$ are seen to converge to the Bondi accretion with time-averaged values of $\approx 9.8$ and $9.5\ M_{\odot}\ {\rm yr}^{-1}$, respectively. Therefore, particle consistency coupled to proper treatment of the outer boundary conditions are important ingredients to ensure convergence to the Bondi accretion rate.

(7) $C^1$-consistency (or second-order accuracy) for the particle approximation is restored when the discrete zeroth-order moment of a function and the first-order moment of the gradient satisfies the kernel consistency constraints $M_0 = 1$ and $\mathbf{M}'_1 = \mathbf{I}$, respectively, where $\mathbf{I}$ is the unit tensor. Our results demonstrate that as the resolution is increased from $N = 63^3$, $n_{\rm neigh} = 3941$ to $N = 200^3$, $n_{\rm neigh} = 22512$ the peaks of the distributions of $M_0$ and $\mathbf{M}'_1$ become so close to unity that approximate $C^1$-consistency is guaranteed. In contrast, the runs with the standard version of the GADGET-2 code produced much broader distributions with $M_0$ and $\mathbf{M}'_1$ peaking at $\approx 0.83$ and $\approx 0.996$, implying even loss of $C^0$-consistency as $M_0 \neq 1$. Due to the symmetry of the kernel the peaks of the distributions of the other moments are always close to zero in all simulations independently of the SPH method and numerical resolution.

(8) In all models with the modified GADGET-2 code the accretion flow remained spherically symmetric, relaxing into a steady flow. This is in contrast with existing simulations of non-stationary accretors where a bow shock is known to form. Conversely, the models with the standard GADGET-2 and GIZMO codes all developed small-scale density and velocity gradients near the accretor as a result of non-linear growth of numerical noise inherent in SPH, which led to different flow patterns as the resolution was increased.

Future studies in this line will include the effects of rotation and growth of the SMBH mass due to non-spherical accretion. Coupling of these dynamical effects to the radiative heating by a central X-ray corona and radiative cooling will point towards the construction of physically more consistent models. Our long-term goal focus on the development of consistent SPH models of AGN feedback, where the radiative-outflow component must be accounted for along with improved microphysics, including the relationships between ionization, molecular states, level populations, and kinetic temperatures of low density environments (Ramírez-Velasquez et al. 2016).

## 7 ACKNOWLEDGMENTS

We thank the anonymous referee who has raised a number of clarifying suggestions that have improved the content of the manuscript. IMPETUS is a collaboration project between the ABACUS-Centro de Matemática Aplicada y Cómputo de Alto Rendimiento of Cinvestav-IPN, the Centro de Física of the Instituto Venezolano de Investigaciones Científicas (IVIC), and the Área de Física de Procesos Irreversibles of the Departamento de Ciencias Básicas of the Universidad Autónoma Metropolitana–Azcapotzalco (UAM-A) aimed at the SPH modelling of astrophysical flows. The project is supported by ABACUS under grant EDOMEX-2011-C01-165873, by IVIC under the project 2013000259, and by UAM-A through internal funds. JMRV thanks the hospitality, support, and computing facilities of ABACUS where this work was done.

**Table 1.** Model simulations of spherical Bondi accretion.

| Run | $r_{\rm out}$ (pc) | $r_{\rm in}$ (pc) | $N$ | $n_{\rm neigh}$ | $M_{\rm tot}$ ($10^7$ $M_\odot$) | $\gamma_{\rm r}$ | $t_{\rm end}$ (kyr) | $\langle \dot{M} \rangle (M_\odot$ yr$^{-1})^a$ |
|---|---|---|---|---|---|---|---|---|
| B1C$_{\rm ISO}$ | 10 | 0.02 | $63^3$ | 64 | 1.0 | 1.0 | 15 | 20.1 |
| B2C$_{\rm ISO}$ | 10 | 0.02 | $80^3$ | 64 | 1.0 | 1.0 | 15 | 18.7 |
| B3C$_{\rm ISO}$ | 10 | 0.02 | $100^3$ | 64 | 1.0 | 1.0 | 15 | 10.1 |
| B4C$_{\rm ISO}$ | 10 | 0.02 | $126^3$ | 64 | 1.0 | 1.0 | 15 | 10.1 |
| B5C$_{\rm ISO}$ | 10 | 0.02 | $159^3$ | 64 | 1.0 | 1.0 | 15 | 9.8 |
| B6C$_{\rm ISO}$ | 10 | 0.02 | $200^3$ | 64 | 1.0 | 1.0 | 15 | 8.9 |
| B1W$_{\rm ISO}$ | 10 | 0.02 | $63^3$ | 3941 | 1.0 | 1.0 | 15 | 162.2 |
| B2W$_{\rm ISO}$ | 10 | 0.02 | $80^3$ | 5651 | 1.0 | 1.0 | 15 | 154.4 |
| B3W$_{\rm ISO}$ | 10 | 0.02 | $100^3$ | 7913 | 1.0 | 1.0 | 15 | 81.6 |
| B4W$_{\rm ISO}$ | 10 | 0.02 | $126^3$ | 11213 | 1.0 | 1.0 | 15 | 81.2 |
| B5W$_{\rm ISO}$ | 10 | 0.02 | $159^3$ | 15927 | 1.0 | 1.0 | 15 | 81.1 |
| B6W$_{\rm ISO}$ | 10 | 0.02 | $200^3$ | 22512 | 1.0 | 1.0 | 15 | 80.6 |
| B4GIZ$_{\rm (CS-ISO)}$ | 10 | 0.02 | $126^3$ | 64 | 1.0 | 1.001 | 15 | 17.0 |
| B4GIZ$_{\rm (WD-ISO)}$ | 10 | 0.02 | $126^3$ | 64 | 1.0 | 1.001 | 15 | 21.1 |
| B4GIZ$_{\rm (FM-ISO)}$ | 10 | 0.02 | $126^3$ | 64 | 1.0 | 1.001 | 15 | 39.9 |
| B6GIZ$_{\rm (CS-ISO)}$ | 10 | 0.02 | $200^3$ | 64 | 1.0 | 1.001 | 15 | 32.1 |
| B6GIZ$_{\rm (WD-ISO)}$ | 10 | 0.02 | $200^3$ | 64 | 1.0 | 1.001 | 15 | 20.3 |
| B6GIZ$_{\rm (FM-ISO)}$ | 10 | 0.02 | $200^3$ | 64 | 1.0 | 1.001 | 15 | 34.7 |
| B1C$_{\rm ADIA}$ | 10 | 0.02 | $63^3$ | 64 | 1.0 | 5/3 | 15 | 4.4 |
| B2C$_{\rm ADIA}$ | 10 | 0.02 | $80^3$ | 64 | 1.0 | 5/3 | 15 | 2.4 |
| B3C$_{\rm ADIA}$ | 10 | 0.02 | $100^3$ | 64 | 1.0 | 5/3 | 15 | 2.6 |
| B4C$_{\rm ADIA}$ | 10 | 0.02 | $126^3$ | 64 | 1.0 | 5/3 | 15 | 1.4 |
| B5C$_{\rm ADIA}$ | 10 | 0.02 | $159^3$ | 64 | 1.0 | 5/3 | 15 | 1.4 |
| B6C$_{\rm ADIA}$ | 10 | 0.02 | $200^3$ | 64 | 1.0 | 5/3 | 15 | 1.4 |
| B1W$_{\rm ADIA}$ | 10 | 0.02 | $63^3$ | 3941 | 1.0 | 5/3 | 15 | 29.2 |
| B2W$_{\rm ADIA}$ | 10 | 0.02 | $80^3$ | 5651 | 1.0 | 5/3 | 15 | 19.1 |
| B3W$_{\rm ADIA}$ | 10 | 0.02 | $100^3$ | 7913 | 1.0 | 5/3 | 15 | 19.6 |
| B4W$_{\rm ADIA}$ | 10 | 0.02 | $126^3$ | 11213 | 1.0 | 5/3 | 15 | 15.8 |
| B5W$_{\rm ADIA}$ | 10 | 0.02 | $159^3$ | 15927 | 1.0 | 5/3 | 15 | 9.8 |
| B6W$_{\rm ADIA}$ | 10 | 0.02 | $100^3$ | 22512 | 1.0 | 5/3 | 15 | 9.5 |

(a) Time averaged mass inflow rate at $r_{in} = 0.02$ pc. Values for the isothermal models are calculated over the entire evolution and for $t \geq 10$ kyr for the adiabatic models.

## REFERENCES

Arreaga-García, G., Klapp, J., Sigalotti, L. Di G., & Gabbasov, R. 2007, ApJ, 666, 290
Balick, B. & Heckman, T. M. 1982, ARAA, 20, 431
Barai, P. 2008, ApJ, 682, L17
Barai, P., Proga, D., & Nagamine, K. 2011, MNRAS, 418, 591
—. 2012, MNRAS, 424, 728
Berger, M. J. & Colella, P. 1989, J. Comput. Phys., 82, 64
Blandford, R. D. 1976, MNRAS, 176, 465
Blandford, R. D. & Begelman, M. C. 1999, MNRAS, 303, L1
—. 2004, MNRAS, 349, 68
Blondin, J. M. & Pope, T. C. 2009, ApJ, 700, 95
Blondin, J. M. & Raymer, E. 2012, ApJ, 752:30 (6pp)
Blondin, J. M., Kallman, T. R., Fryxell, B. A., & Taam, R. E. 1990, ApJ, 356, 591
Bondi, H. 1952, MNRAS, 112, 195
Boss, A. P. 1989, ApJ, 345, 554
Bürzle, F., Clark, P. C., Stasyszyn, F., Greif, T., Dolag, K., Klessen, R. S., & Nielaba, P. 2011, MNRAS, 412, 171
Braito, V., Reeves, J. N., Dewangan, G. C., George, I., Griffiths, R. E., Markowitz, A., Nandra, K., Porquet, D., Ptak, A., Turner, T. J., Yaqoob, T., & Weaver, K. 2007, ApJ, 670, 978
Ciotti, L. & Ostriker, J. P. 2001, ApJ, 551, 131
Commerçon, B., Hannebelle, P., Audit, E., Chabrier, G., & Teyssier, R. 2008, A&A, 482, 371
Couchman, H. M. P., Thomas, P. A., & Pearce, F. R. 1995, ApJ, 452, 797
Cowie, L. L. 1977, MNRAS, 180, 491
Cullen, L. & Dehnen, W. 2010, MNRAS, 408, 669
Davies, R. E. & Pringle, J. E. 1980, MNRAS, 191, 599
Dehnen, W. & Aly, H. 2012, MNRAS, 425, 1068
Dewangan, G. C., Griffiths, R. E., Dasgupta, S., & Rao, A. R. 2007, ApJ, 671, 1284
Dewangan, G. C., Jithesh, V., Misra, R., & Ravikumar, C. D. 2013, ApJ, 771, L37
Di Matteo, T., Springel, V., & Hernquist, L. 2005, Nature, 433, 604
Edgar, R. 2004, New Astron. Rev., 48, 843
Fabian, A. C. 1999, MNRAS, 308, L39
Fabian, A. C. & Crawford, C. S. 1990, MNRAS, 247, 439
Farrell, S. A., Webb, N. A., Barret, D., Godet, O., & Rodrigues, J. M. 2009, Nature, 460, 73
Faucher-Giguère, C.-A. & Quataert, E. 2012, MNRAS, 425, 605
Ferrarese, L. & Ford, H. 2005, Space Sci. Rev., 116, 523
Foglizzo, T. 2001, A&A, 368, 311
Foglizzo, T. & Tagger, M. 2000, A&A, 363, 174
Foglizzo, T., Galletti, P., & Ruffert, M. 2005, A&A, 435, 397
Font, J. A. & Ibáñez, J. M. A. 1998, ApJ, 494, 297
Fryxell, B. A., Taam, R. E., & McMillan, S. L. W. 1987, ApJ, 315, 536
Fukue, J. 2001, PASJ, 53, 687
Gabbasov, R., Sigalotti, L. Di G., Cruz, F., Klapp, J., & Ramírez-Velasquez, J. M. 2017, ApJ, 835, id. 287 (25pp)
Gebhardt, K., Bender, R., Bower, G., Dressler, A., Faber, S. M., Filippenko, A. V., Green, R., Grillmair, C., Ho, L. C., Kormendy, J., Lauer, T. R., Magorrian, J., Pinkney, J., Richstone, D., & Tremaine, S. 2000, ApJ, 539, L13
Germain, J., Barai, P., & Martel, H. 2009, ApJ, 704, 1002
Hayes, J. C., Norman, M. L., Fiedler, R. A., Bordner, J. O., Li, P. S., Clark, S. E., ud-Doula, A., & Mac Low, M.-M. 2006, ApJS, 165, 188
Hopkins, P. F. 2015, MNRAS, 450, 53
Hu, C.-Y., Naab, T., Walch, S., Moster, B. P., & Oser, L. 2014, MNRAS, 443, 1173
Kauffmann, G. & Haehnelt, M. 2000, MNRAS, 311, 576
King, A. & Lasota, J.-P. 2016, MNRAS, 458, L10
Koide, H., Matsuda, T., & Shima, E. 1991, MNRAS, 252, 473
Kormendy, J. & Richstone, D. 1995, ARAA, 33, 581
Kovalenko, I. G. & Eremin, M. A. 1998, MNRAS, 298, 861
Krumholz, M. R., McKee, C. F., & Klein, R. I. 2005, ApJ, 618, 757
Kurosawa, R., Proga, D., & Nagamine, K. 2009, ApJ, 707, 823
Li, Y., Hernquist, L., Robertson, B., Cox, T. J., Hopkins, P. F., Springel, V., Gao, L., Di Matteo, T., Zentner, A. R., Jenkins, A., & Yoshida, N. 2007, ApJ, 665, 187
Livio, M., Soker, N., Matsuda, T., & Anzer, U. 1991, MNRAS, 253, 633
Luangtip, W., Roberts, T. P., & Done, C. 2016, MNRAS, 460, 4417
Lynden-Bell, D. 1969, Nature, 223, 690
Magorrian, J., Tremaine, S., Richstone, D., Bender, R., Bower, G., Dressler, A., Faber, S. M., Gebhardt, K., Green, R., Grillmair, C., Kormendy, J., & Lauer, T. 1998, AJ, 115, 2285
Matsuda, T., Inoue, M., & Sawada, K. 1987, MNRAS, 226, 785
Matsuda, T., Sekino, N., Sawada, K., Shima, E., Livio, M., Anzer, U., & Börner, G. 1991, A&A, 248, 301
Miller, M. C. & Colbert, E. J. M. 2004, Int. J. Modern Phys. D, 13, 1
Moeckel, N. & Throop, H. B. 2009, ApJ, 707, 268
Mościbrodzka, M. & Proga, D. 2013, ApJ, 767, 156
Nagamine, K., Barai, P., & Proga, D. 2012, in AGN Winds in Charleston (ASP Conference Series, Vol. 460), ed. G. Chartas, F. Hamann, & K. M. Leighly (San Francisco: ASP), 245
Narayan, R., Sądowski, A., & Soria, R. 2017, MNRAS, 469, 2997
Novak, G. S., Ostriker, J. P., & Ciotti, L. 2011, ApJ, 737, 26
Orosz, J. A. 2003, in A Massive Star Odyssey: From Main Sequence to Supernova (Proceedings of IAU Symposium No. 212), ed. K. A. van der Hucht, A. Herrero, & C. Esteban (San Francisco: ASP), 365
Ostriker, J. P., Choi, E., Ciotti, L., Novak, G. S., & Proga, D. 2010, ApJ, 722, 642
Ozernoi, L. M. & Reinhardt, M. 1978, Ap&SS, 59, 171
Paczyńsky, B. & Wiita, P. J. 1980, A&A, 88, 23
Perna, R., Narayan, R., Rybicki, G., Stella, L., & Treves, A. 2003, ApJ, 594, 936
Pogorelov, N. V., Ohsugi, Y., & Matsuda, T. 2000, MNRAS, 313, 198
Proga, D. 2000, ApJ, 538, 684
—. 2003, ApJ, 592, L9
—. 2007, ApJ, 661, 693
Proga, D. & Begelman, M. C. 2003, ApJ, 582, 69
Proga, D. & Kallman, T. R. 2004, ApJ, 616, 688
Proga, D., Stone, J. M., & Kallman, T. R. 2000, ApJ, 543, 686
Quataert, E. & Gruzinov, A. 2000, ApJ, 539, 809
Ramírez, J. M. 2008, A&A, 489, 57
—. 2013, A&A, 551, A95
Ramírez-Velasquez, J. M., Klapp, J., Gabbasov, R., Cruz, F., & Sigalotti, L. D. G. 2016, ApJS, 226, id. 22 (13pp)
Rasio, F. A. 2000, Prog. Theoret. Phys. Suppl., 138, 609
Read, J. I., Hayfield, T., & Agertz, O. 2010, MNRAS, 405, 1513
Rees, M. J. 1984, ARAA, 22, 471
Riaz, R., Farooqui, S. Z., & Vanaverbeke, S. 2014, MNRAS, 444, 1189
Ricotti, M. 2007, ApJ, 662, 53
Riebe, K., Partl, A. M., Enke, H., Forero-Romero, J., Gottlöber, S., Klypin, A., Lemson, G., Prada, F., Primack, J. R., Steinmetz, M., & Turchaninov, V. 2013, Astronomische Nachrichten, 334, 691
Robinet, J. C., Gressier, J., Casalis, G., & Moschetta, J. M. 2000, J. Fluid Mech., 417, 237
Rosswog, S. 2015, Living Rev. Comput. Astrophys., 1, 1
Ruffert, M. 1995, A&AS, 113, 133
Ruffert, M. 1996, A&A, 311, 817
Ruffert, M. 1997, A&A, 317, 793
Ruffert, M. 1999, A&A, 346, 861

Ruffert, M. & Arnett, D. 1994, ApJ, 427, 351
Ruffert, M. & Anzer, U. 1995, A&A, 295, 108
Russell, H. R., Fabian, A. C., McNamara, B. R., & Broderick, A. E. 2015, MNRAS, 451, 588
Salpeter, E. E. 1964, ApJ, 140, 796
Sari, R. & Goldreich, P. 2006, ApJ, 642, L65
Sawada, K., Matsuda, T., Anzer, U., Börner, G., & Livio, M. 1989, A&A, 221, 263
Shima, E., Matsuda, T., Takeda, H., & Sawada, K. 1985, MNRAS, 217, 367
Shima, E., Matsuda, T., Anzer, U., Börner, G., & Boffin, H. M. J. 1998, A&A, 337, 311
Sigalotti, L. Di G. 1994, A&A, 283, 858
Sigalotti, L. Di G., Klapp, J., Rendón, O., Vargas, C. A., & Peña-Polo, F. 2016a, Appl. Num. Math., 108, 242
Sigalotti, L. Di G., Rendón, O., Klapp, J., Vargas, C. A., & Campos, K. 2016b, arXiv:1608.05883v1 [physics.comp-ph]
Soker, N. 1990, ApJ, 358, 545
Soker, N. 1991, ApJ, 376, 750
Springel, V. 2005, MNRAS, 364, 1105
Springel, V. & Hernquist, L. 2002, MNRAS, 333, 649
Springel, V., White, S. D. M., Jenkins, A., Frenk, C. S., Yoshida, N., Gao, L., Navarro, J., Thacker, R., Croton, D., Helly, J., Peacock, J. A., Cole, S., Thomas, P., Couchman, H., Evrard, A., Colberg, J., & Pearce, F. 2005, Nature, 435, 629
Taam, R. E., Fu, A. & Fryxell, B. A. 1991, ApJ, 371, 696
Wendland, H. 1995, Advances in Computational Mathematics, 4, 389
Wong, K.-W., Irwin, J. A., Yukita, M., Million, E. T., Mathews, W. G. & Bregman, J. N. 2011, ApJ, 736, id. L23 (5pp)
Zhu, Q., Hernquist, L., & Li, Y. 2015, ApJ, 800, id. 6 (13pp)

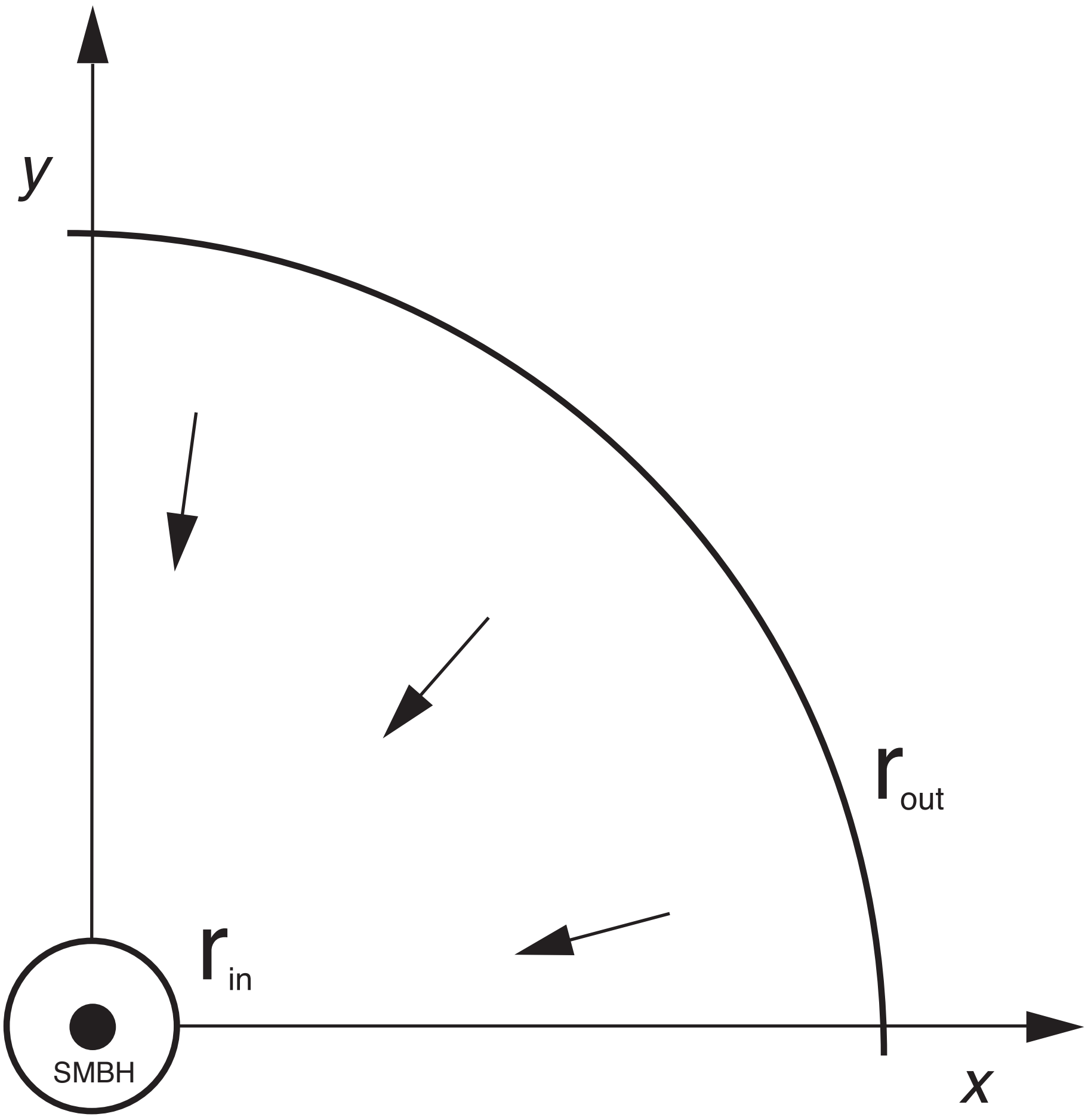

**Figure 1.** Schematic cross-section of the spherical computational domain in the $x$-$y$ plane of a Cartesian coordinate system. The supermassive black hole (SMBH) is located at the origin $x = y = z = 0$. The simulation volume has inner radius $r_{\rm in} = 0.02$ pc and outer radius $r_{\rm out} = 10$ pc.

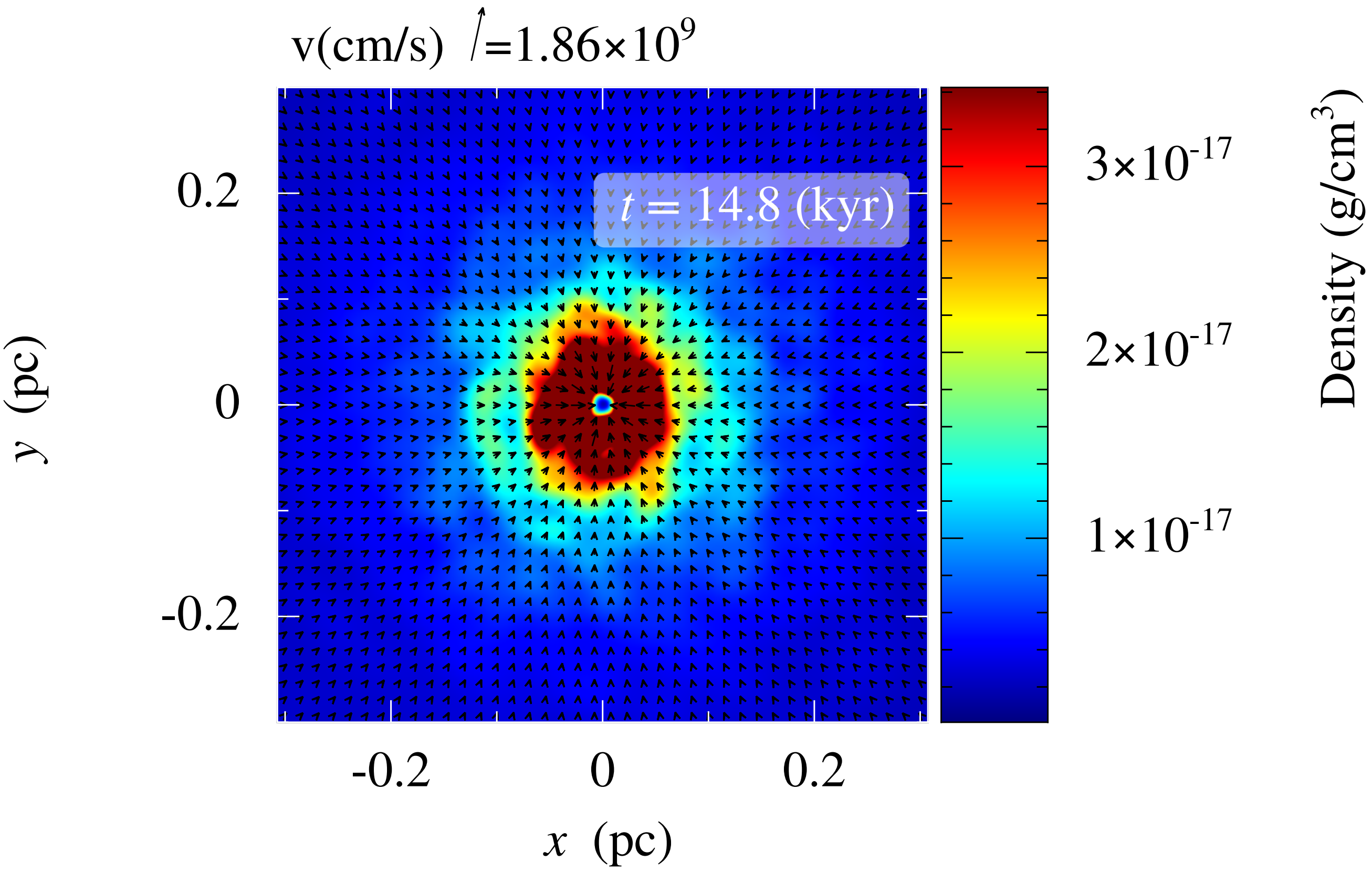

**Figure 2.** Gas density map and velocity vectors in the $x$-$y$ plane near the accretor at 14.8 kyr during the spherical accretion of the highly resolved model B6C$_{\rm ISO}$ working with $N = 200^3$ and $n_{\rm neigh} = 64$. The colour bar and numbers on the right side indicate the density contrasts. On the left top, the scale of the velocity vectors is indicated by the inclined arrow and its magnitude is given in cm s$^{-1}$.

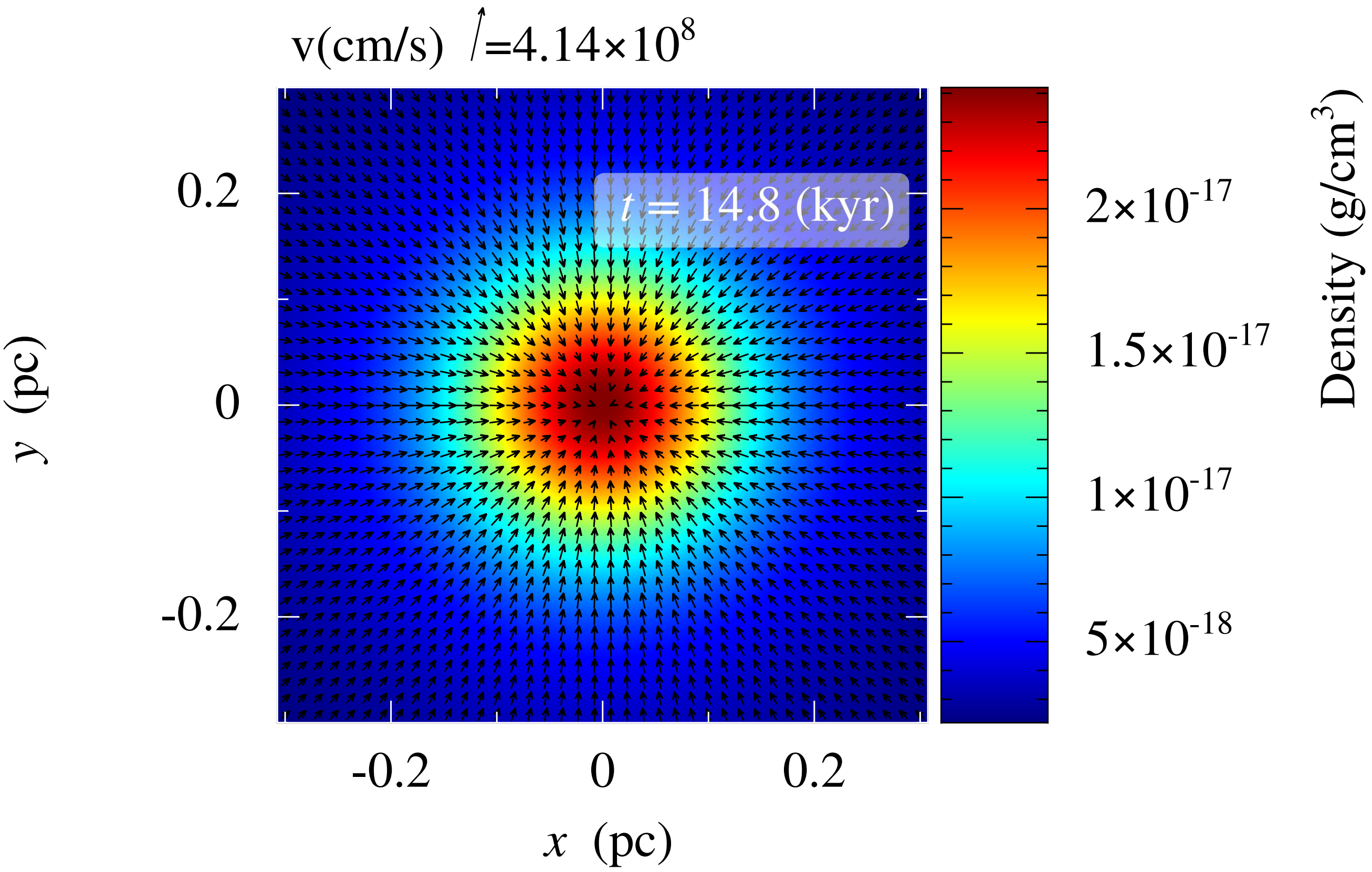

**Figure 3.** Gas density map and velocity vectors in the $x$-$y$ plane near the accretor at 14.8 kyr during the spherical accretion of the highly resolved model B6W$_{\rm ISO}$ working with $N = 200^3$ and $n_{\rm neigh} = 22512$. The colour bar and numbers on the right side indicate the density contrasts. On the left top, the scale of the velocity vectors is indicated by the inclined arrow and its magnitude is given in cm s$^{-1}$.

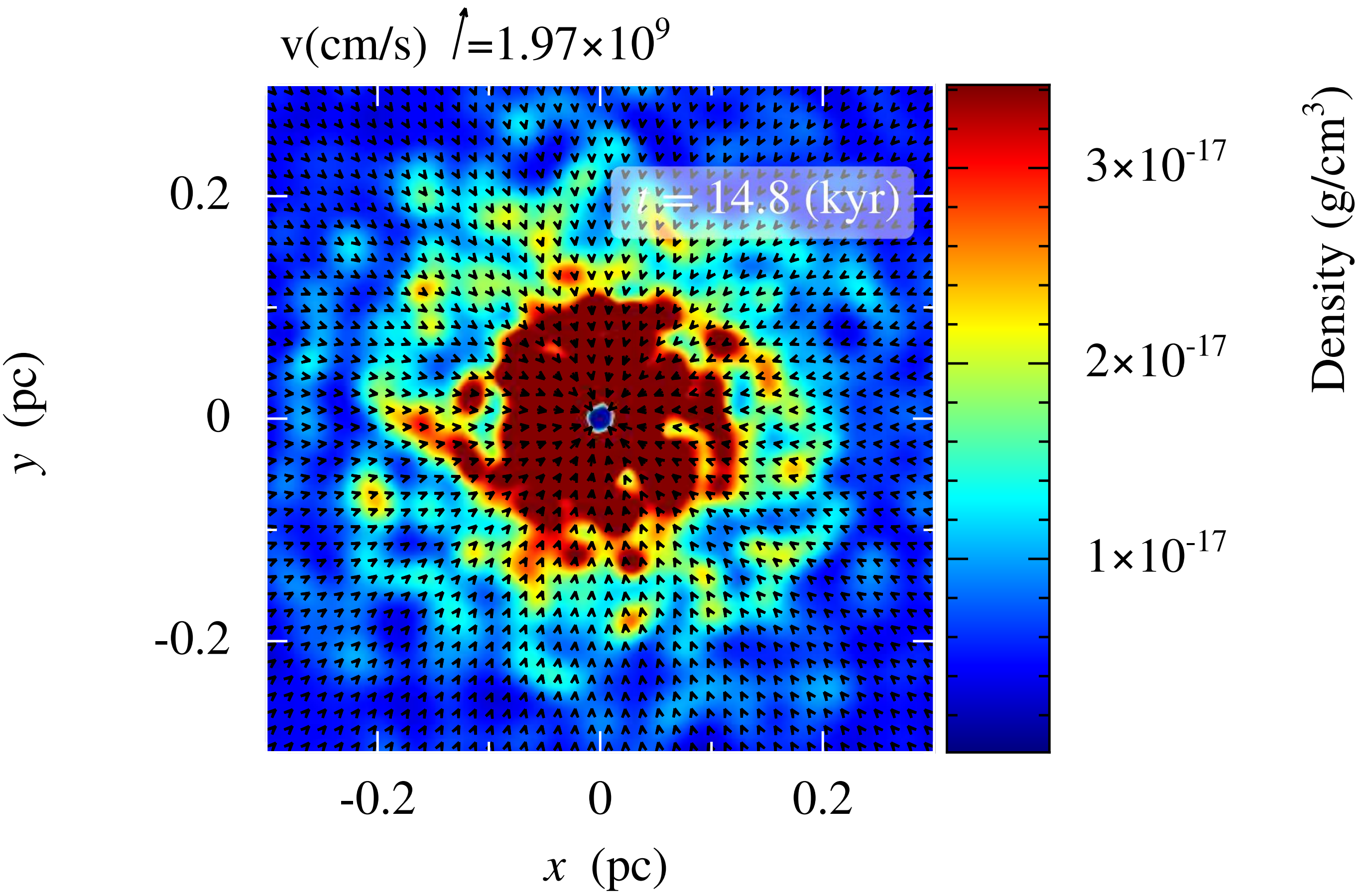

**Figure 4.** Gas density map and velocity vectors in the $x$-$y$ plane near the accretor at 14.8 kyr during the spherical accretion of the highly resolved model B6GIZ$_{(FM-ISO)}$ working with $N = 200^3$ and $n_{neigh} = 64$. The colour bar and numbers on the right side indicate the density contrasts. On the left top, the scale of the velocity vectors is indicated by the inclined arrow and its magnitude is given in cm s$^{-1}$.

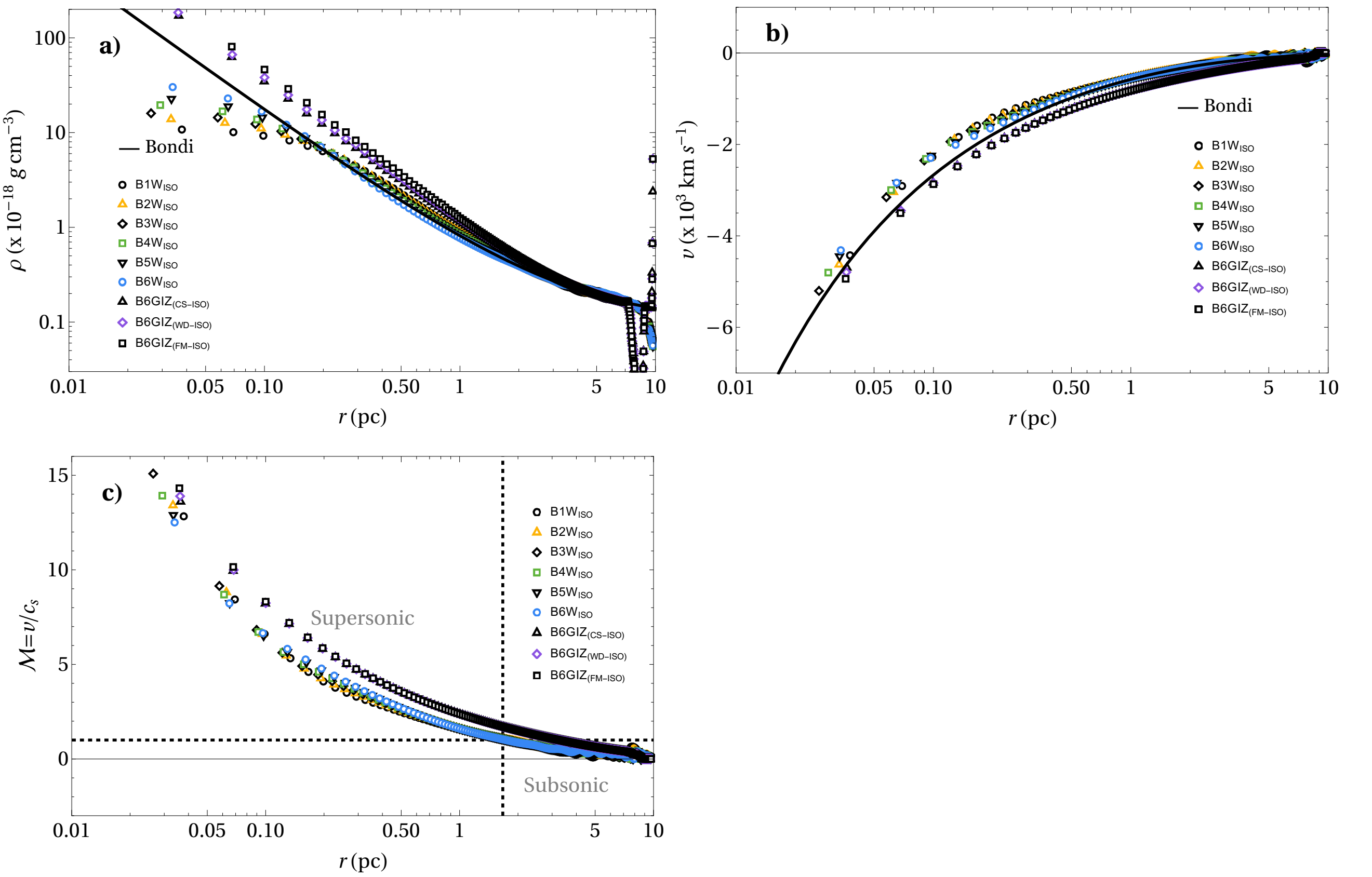

**Figure 5.** a) Density ($\rho$), b) radial velocity ($v$), and c) Mach number ($\mathcal{M} = v/c_s$) profiles at 15 kyr during the spherical accretion for models B1W$_{\rm ISO}$–B6W$_{\rm ISO}$ working with varying number of neighbours as compared with the GIZMO models B6GIZ$_{\rm (CS-ISO)}$, B6GIZ$_{\rm (WD-ISO)}$, and B6GIZ$_{\rm (FM-ISO)}$. The symbols that depict the numerical solutions correspond to the median of the distributions. In a) and b) the solid black line is the exact Bondi solution. The horizontal and vertical dashed lines in c) mark the transition from subsonic to supersonic inflow and the numerically predicted location of the sonic radius $R_s \approx 1.67$ pc, respectively, for models B1W$_{\rm ISO}$–B6W$_{\rm ISO}$.

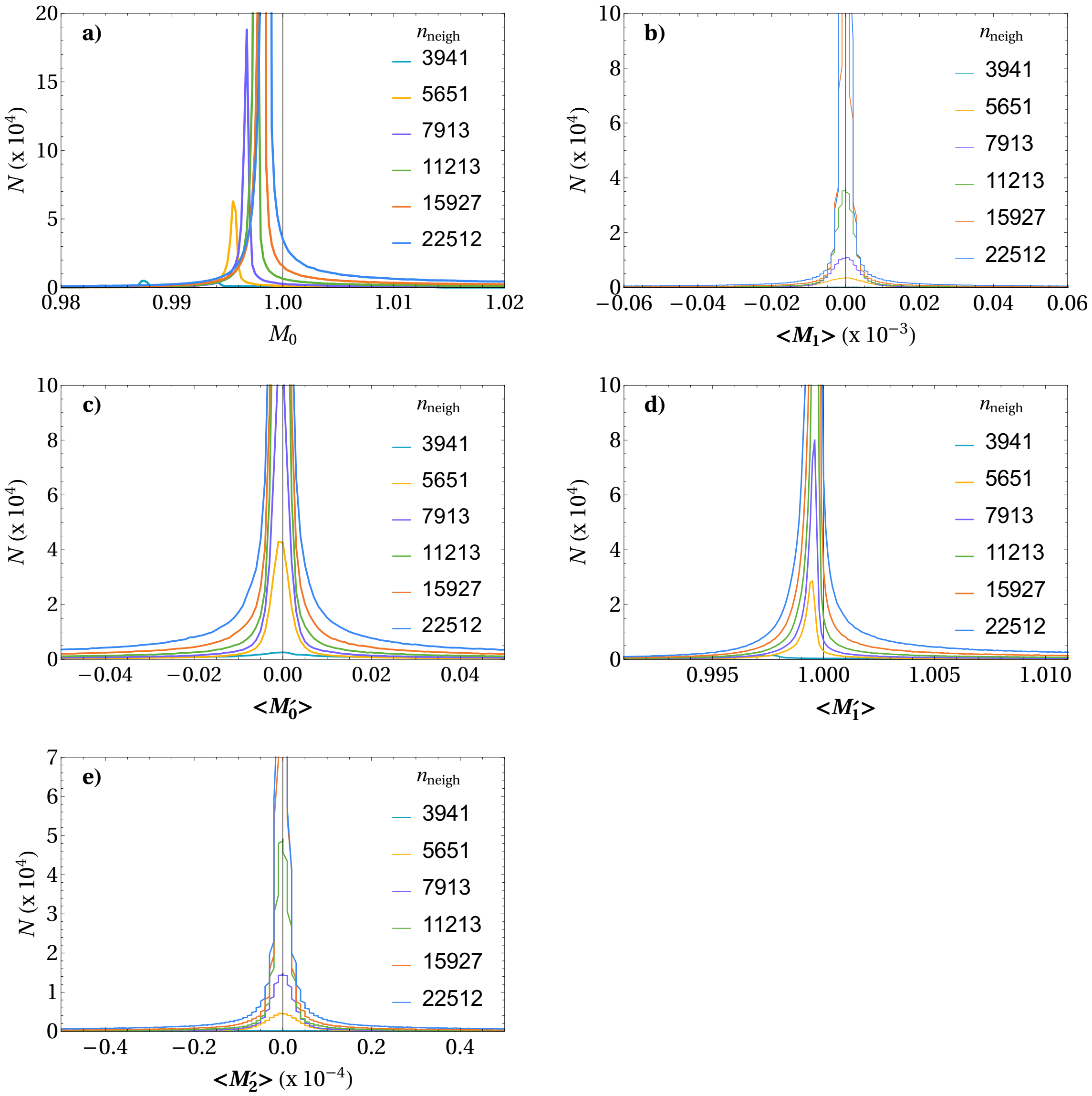

**Figure 6.** Distributions of the moments of the kernel $M_0$ and $\mathbf{M}_1$ as given by Eqs. (13) and (28) (upper plots) and the kernel gradient $\mathbf{M}'_0$, $\mathbf{M}'_1$, and $\mathbf{M}'_2$ as given by Eqs. (29), (14), and (30) (middle and bottom plots), respectively, for models B1W$_{\rm ISO}$–B6W$_{\rm ISO}$. $\langle \mathbf{M}_1 \rangle$ and $\langle \mathbf{M}'_0 \rangle$ are the average values of the components of vectors $\mathbf{M}_1$ and $\mathbf{M}'_0$, while $\langle \mathbf{M}'_1 \rangle$ and $\langle \mathbf{M}'_2 \rangle$ correspond to the average of the components of tensors $\mathbf{M}'_1$ and $\mathbf{M}'_2$, respectively. The vertical solid line in each frame marks the value of the kernel moments for which $C^1$-consistency is achieved.

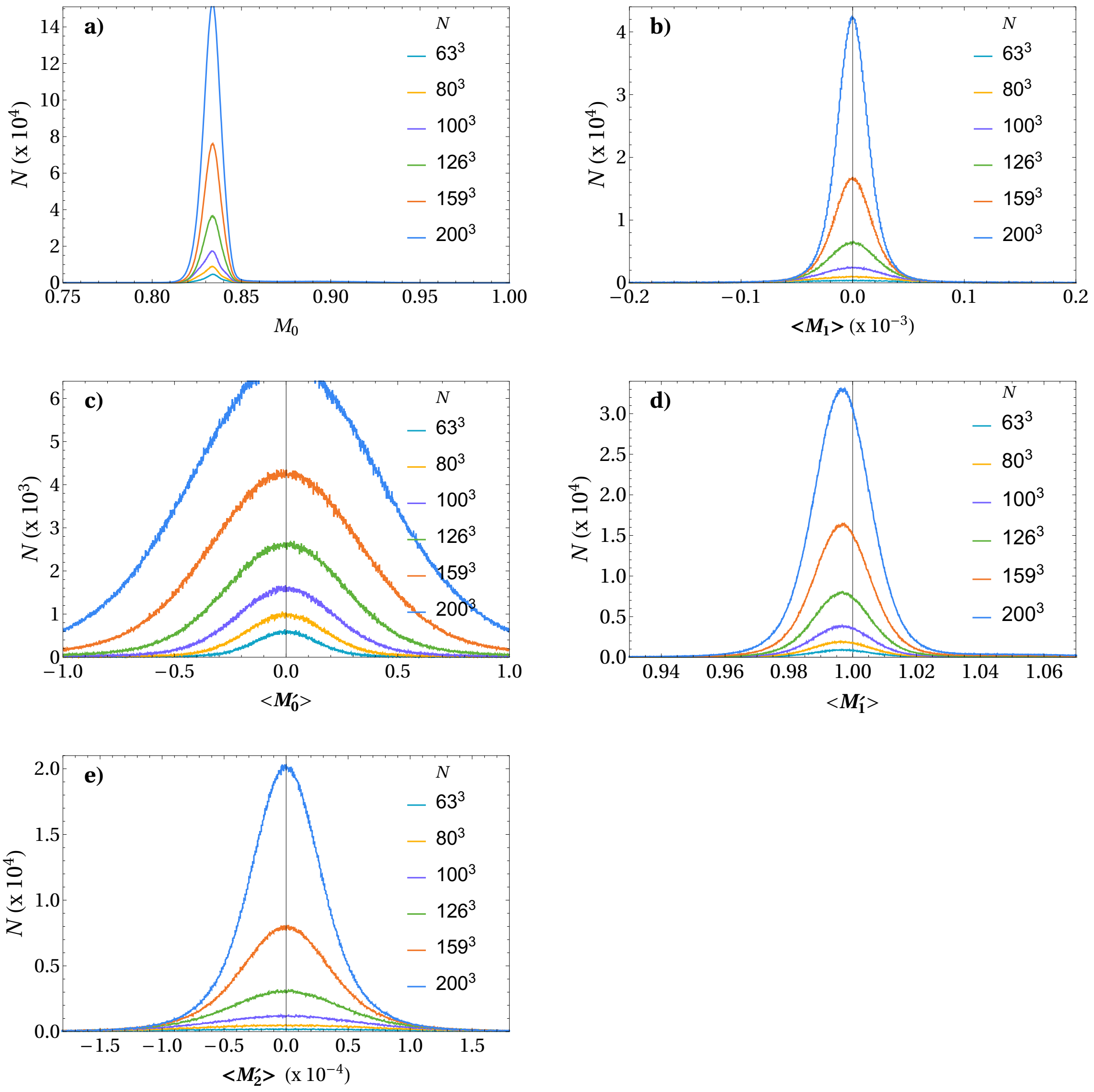

**Figure 7.** The same as Fig. 6 for models B1C$_{\rm ISO}$–B6C$_{\rm ISO}$.

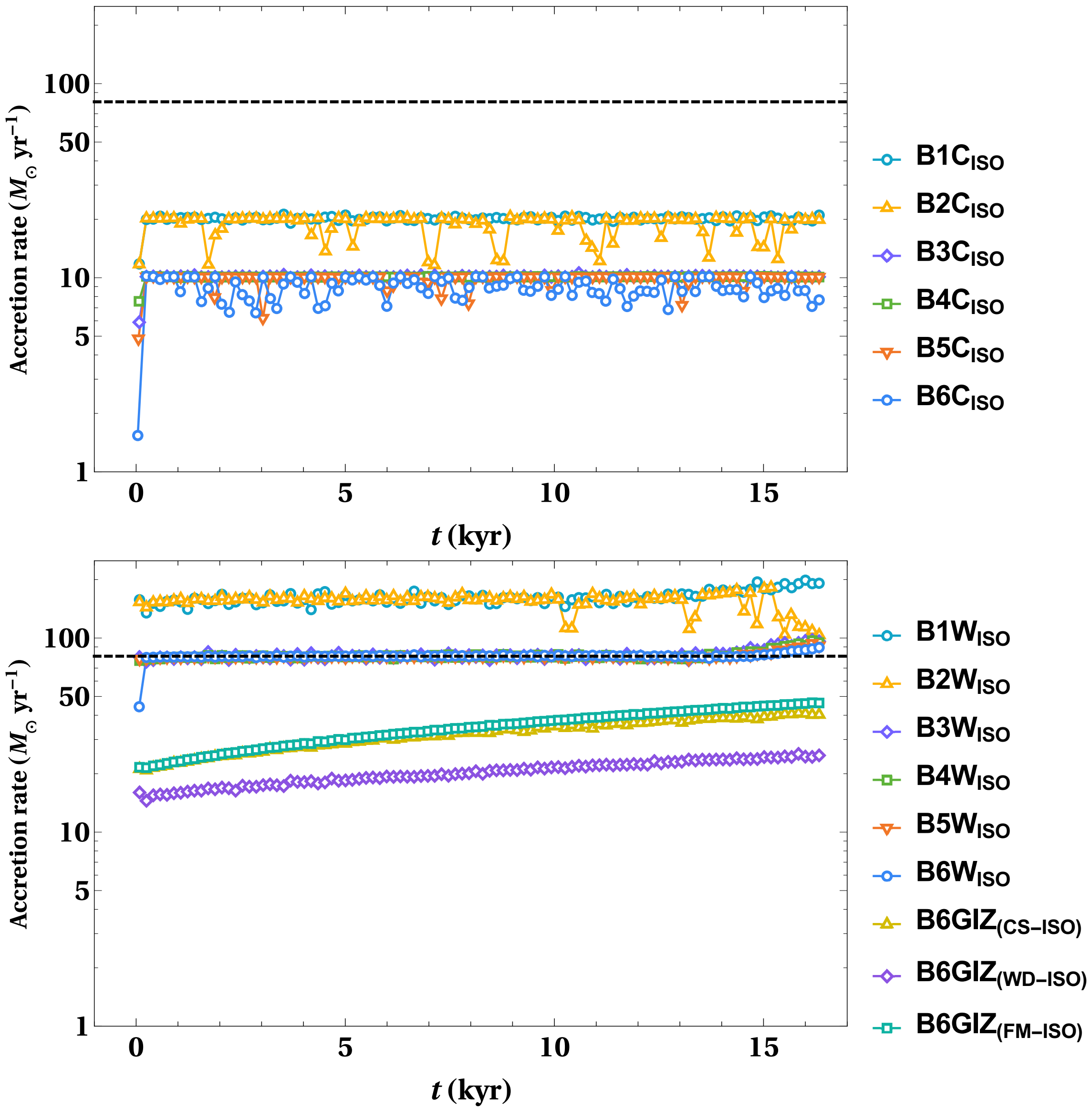

**Figure 8.** Mass inflow rate at $r_{\rm in} = 0.02$ pc for models B1C$_{\rm ISO}$–B6C$_{\rm ISO}$ (top panel) and B1W$_{\rm ISO}$–B6W$_{\rm ISO}$ (bottom panel) as a function of time. Also shown in the bottom panel are the resulting mass inflow rates for the GIZMO models B6GIZ$_{\rm (CS-ISO)}$, B6GIZ$_{\rm (WD-ISO)}$, and B6GIZ$_{\rm (FM-ISO)}$. The horizontal dashed line marks the Bondi accretion rate $\dot{M}_{\rm B} \approx 80.8\ M_\odot$ yr$^{-1}$. A steady-state accretion is quickly achieved by all cases with the exception of the GIZMO models, where the mass inflow rates grow slowly with time. Models B1C$_{\rm ISO}$–B6C$_{\rm ISO}$ predict accretion rates that are factors from $\approx 4$ to 8 times lower than the Bondi rate. In contrast, models B1W$_{\rm ISO}$ and B2W$_{\rm ISO}$ working at lower resolution overestimate the Bondi rate by factors of $\approx 2$, while only the more consistent models B3W$_{\rm ISO}$–B6W$_{\rm ISO}$ closely match the exact value. For comparison, the GIZMO models B6GIZ$_{\rm (CS-ISO)}$ and B6GIZ$_{\rm (FM-ISO)}$ are below the Bondi rate by factors ranging from $\approx 2$ to 4, while model B6GIZ$_{\rm (WD-ISO)}$ exhibits accretion rates that are also lower than the Bondi accretion by factors from $\approx 5$ to 10.

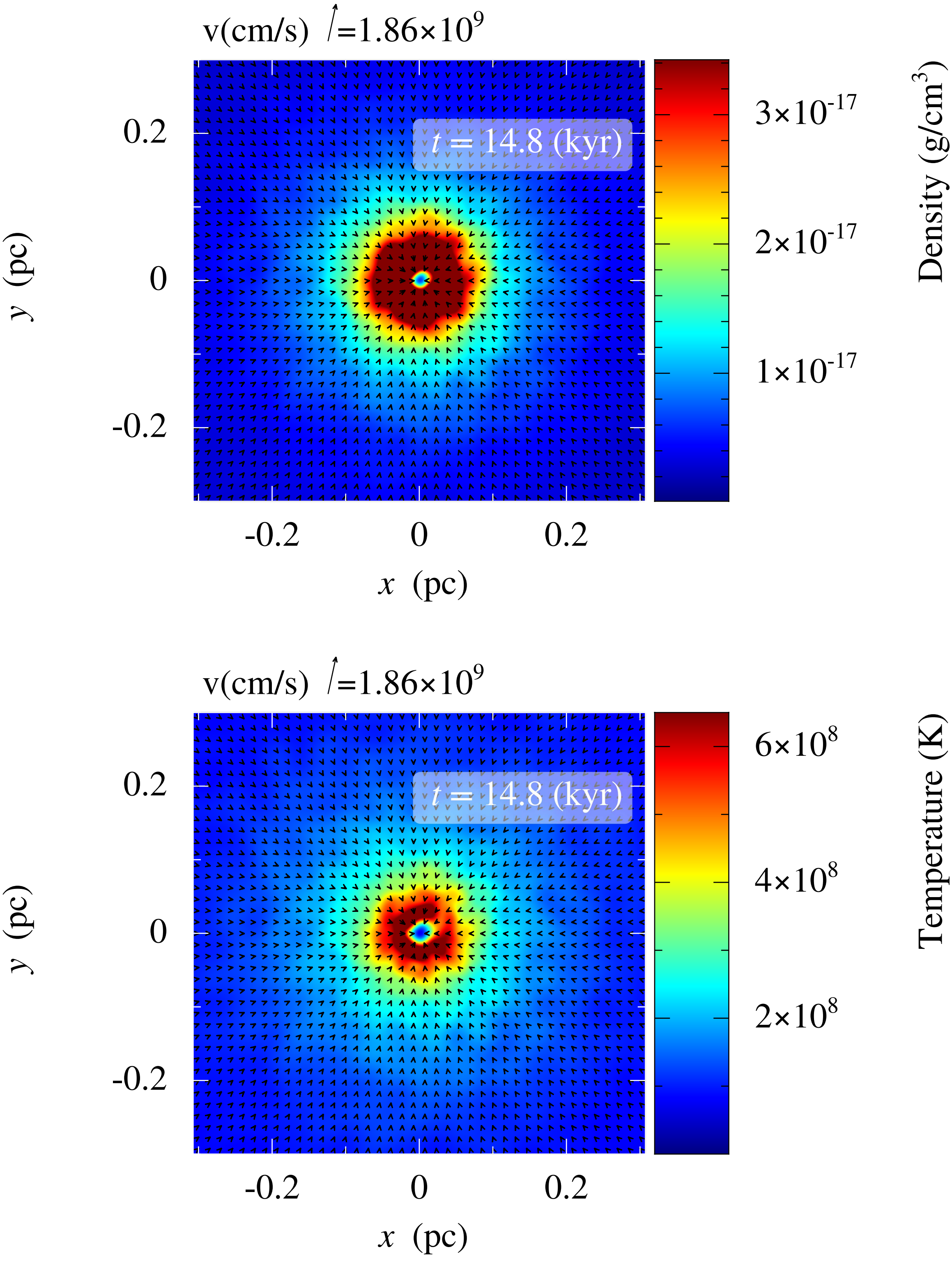

**Figure 9.** Gas density (top) and temperature (bottom) maps overplotted with velocity vectors in the $x$-$y$ plane near the accretor at 14.8 kyr during the spherical accretion of the highly resolved model B6C$_{\rm ADIA}$ working with $N = 200^3$ and $n_{\rm neigh} = 64$. The colour bars and numbers on the right side of the top and bottom plots indicate the density and temperature contrasts, respectively. On the left top of both plots, the scale of the velocity vectors is indicated by the inclined arrow and its magnitude is given in cm s$^{-1}$.

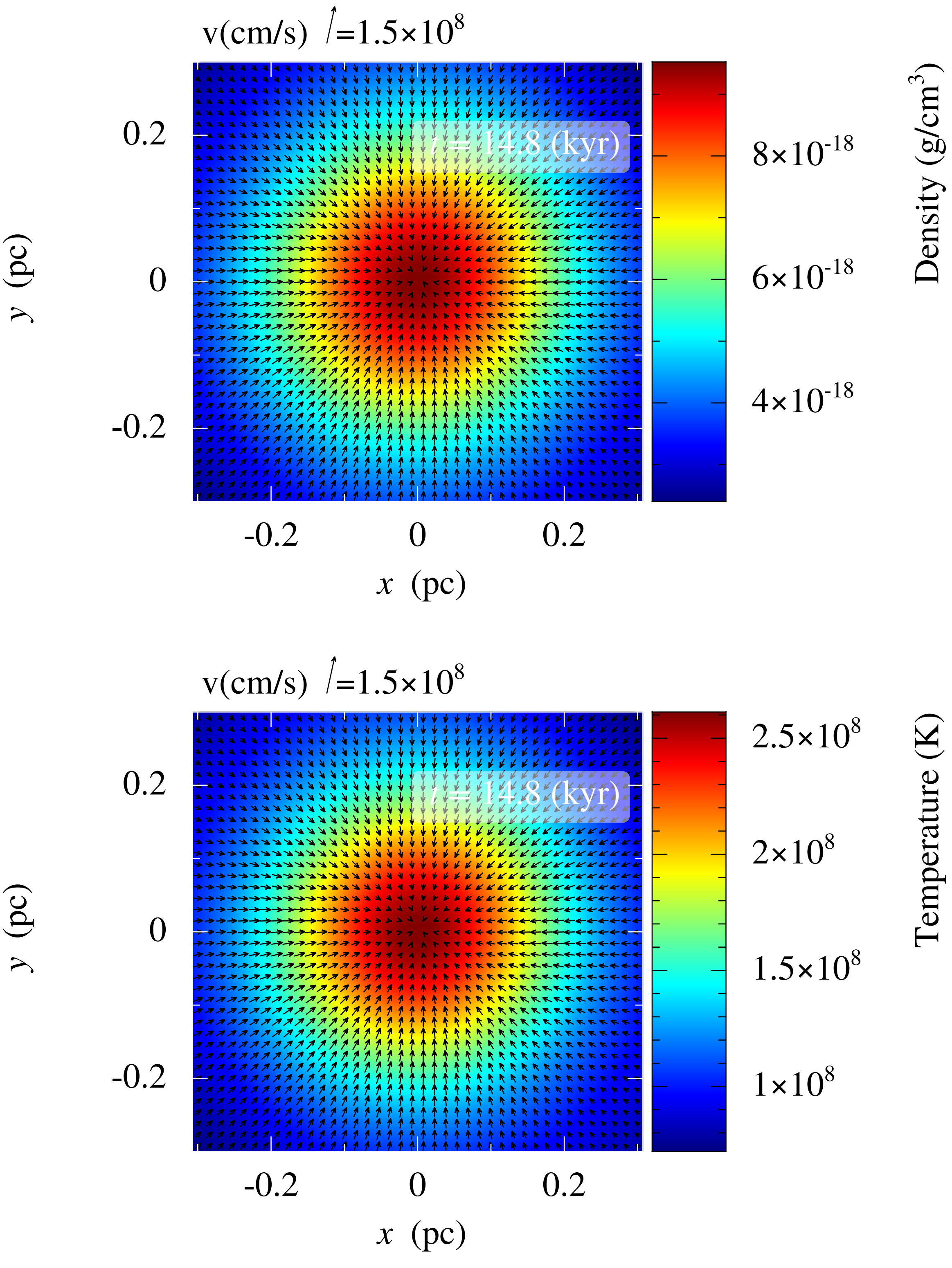

**Figure 10.** Gas density (top) and temperature (bottom) maps overplotted with velocity vectors in the $x$-$y$ plane near the accretor at 14.8 kyr during the spherical accretion of the highly resolved model B6W$_{\rm ADIA}$ working with $N = 200^3$ and $n_{\rm neigh} = 22512$. The colour bars and numbers on the right side of the top and bottom plots indicate the density and temperature contrasts, respectively. On the left top of both plots, the scale of the velocity vectors is indicated by the inclined arrow and its magnitude is given in cm s$^{-1}$.

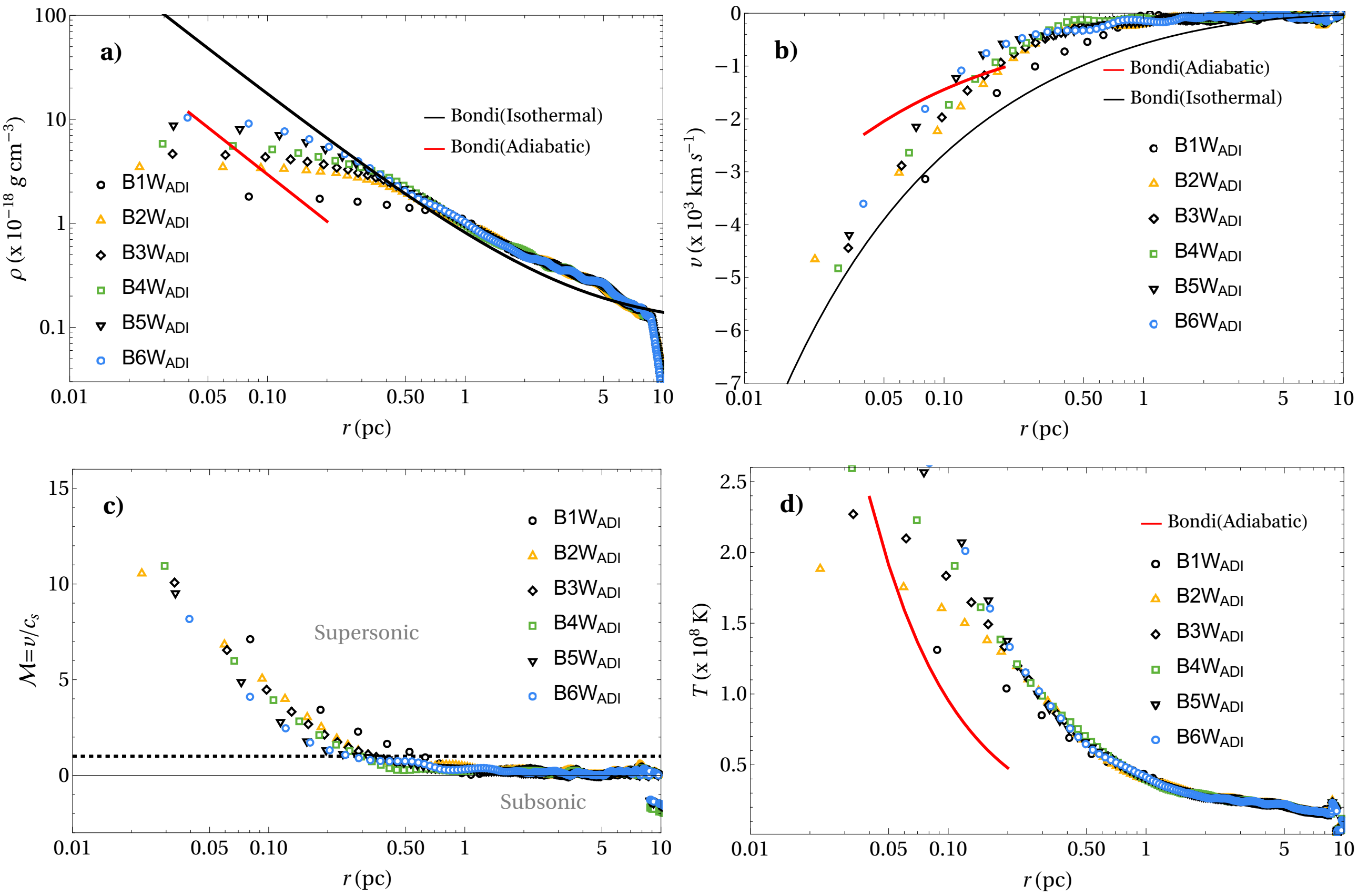

**Figure 11.** a) Density ($\rho$), b) radial velocity ($v$), c) Mach number ($\mathcal{M} = v/c_s$), and d) temperature ($T$) profiles at 15 kyr during the spherical accretion of models B1W$_{ADIA}$–B6W$_{ADIA}$ working with varying number of neighbours. The solid black lines in a) and b) depict the isothermal Bondi solution. The solid red lines in a), b), and d) correspond to limiting forms near the inner boundary as defined by expressions (32)-(34) for the density, radial velocity, and temperature for the adiabatic ($\gamma_r = 5/3$) Bondi accretion. The symbols depict the numerical solutions and correspond to the median of the distributions. The horizontal dashed line in c) marks the transition from subsonic to supersonic inflow.

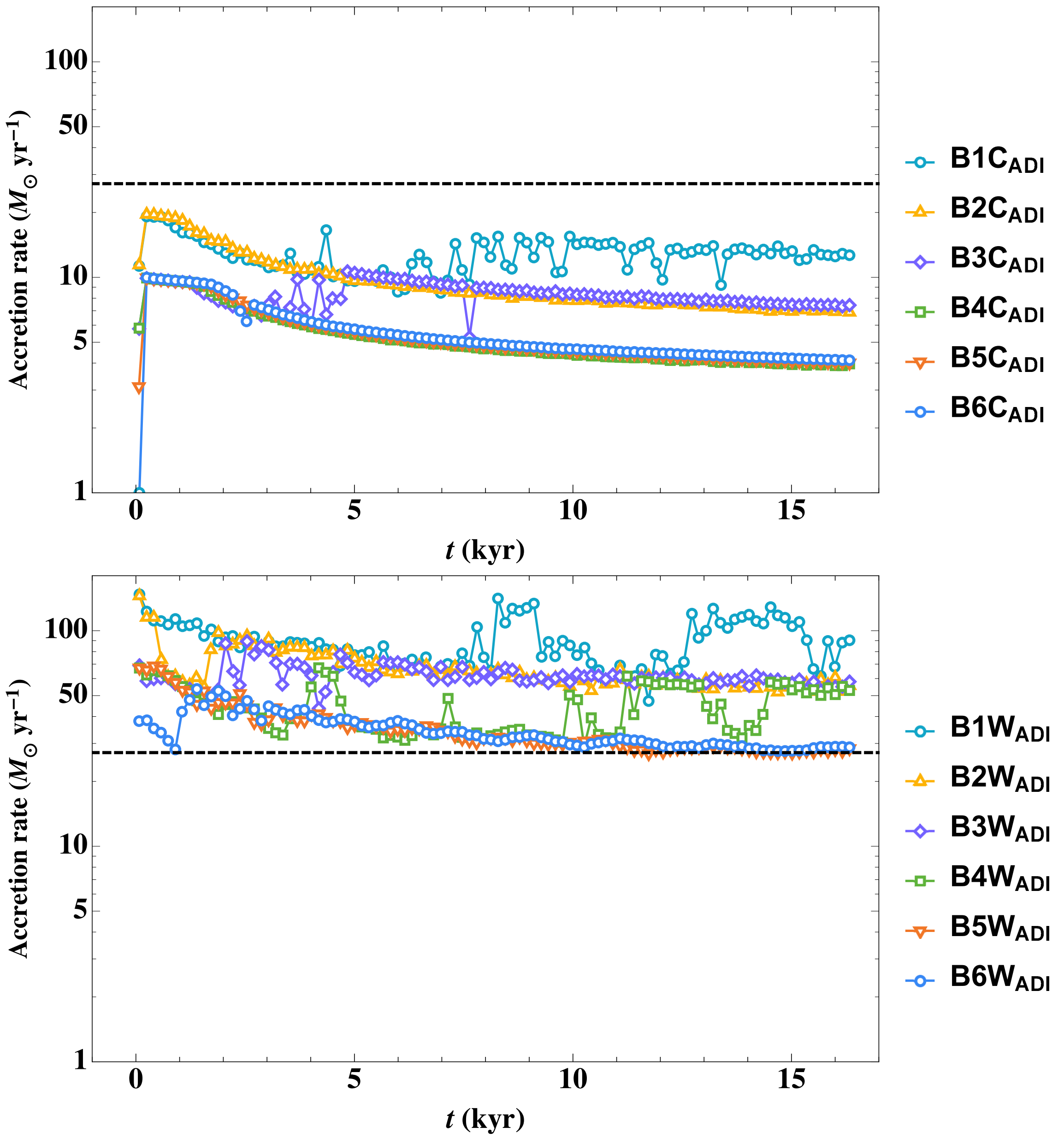

**Figure 12.** Mass inflow rate at $r_{\rm in} = 0.02$ pc for models B1C$_{\rm ADIA}$–B6C$_{\rm ADIA}$ (top panel) and B1W$_{\rm ADIA}$–B6W$_{\rm ADIA}$ (bottom panel) as a function of time. The horizontal dashed line marks the Bondi accretion rate $\dot{M}_{\rm B} \approx 8.6\ M_\odot\ {\rm yr}^{-1}$. Steady-state accretion is approximately reached by both sets of models after $\approx 10$ kyr, except for the lowest resolution runs where considerable scatter is present in the data. Models B1C$_{\rm ADIA}$–B6C$_{\rm ADIA}$ reach steady-state accretion at rates that are much below the Bondi accretion. In contrast, models B1W$_{\rm ADIA}$–B4W$_{\rm ADIA}$ reach steady-state accretion rates that are factor from $\approx 3$ to 2 higher than the Bondi accretion, while the more consistent models B5W$_{\rm ADIA}$ and B6W$_{\rm ADIA}$ are closely approaching the exact solution with values around 9.8 and 9.5 $M_\odot$ yr$^{-1}$, respectively.

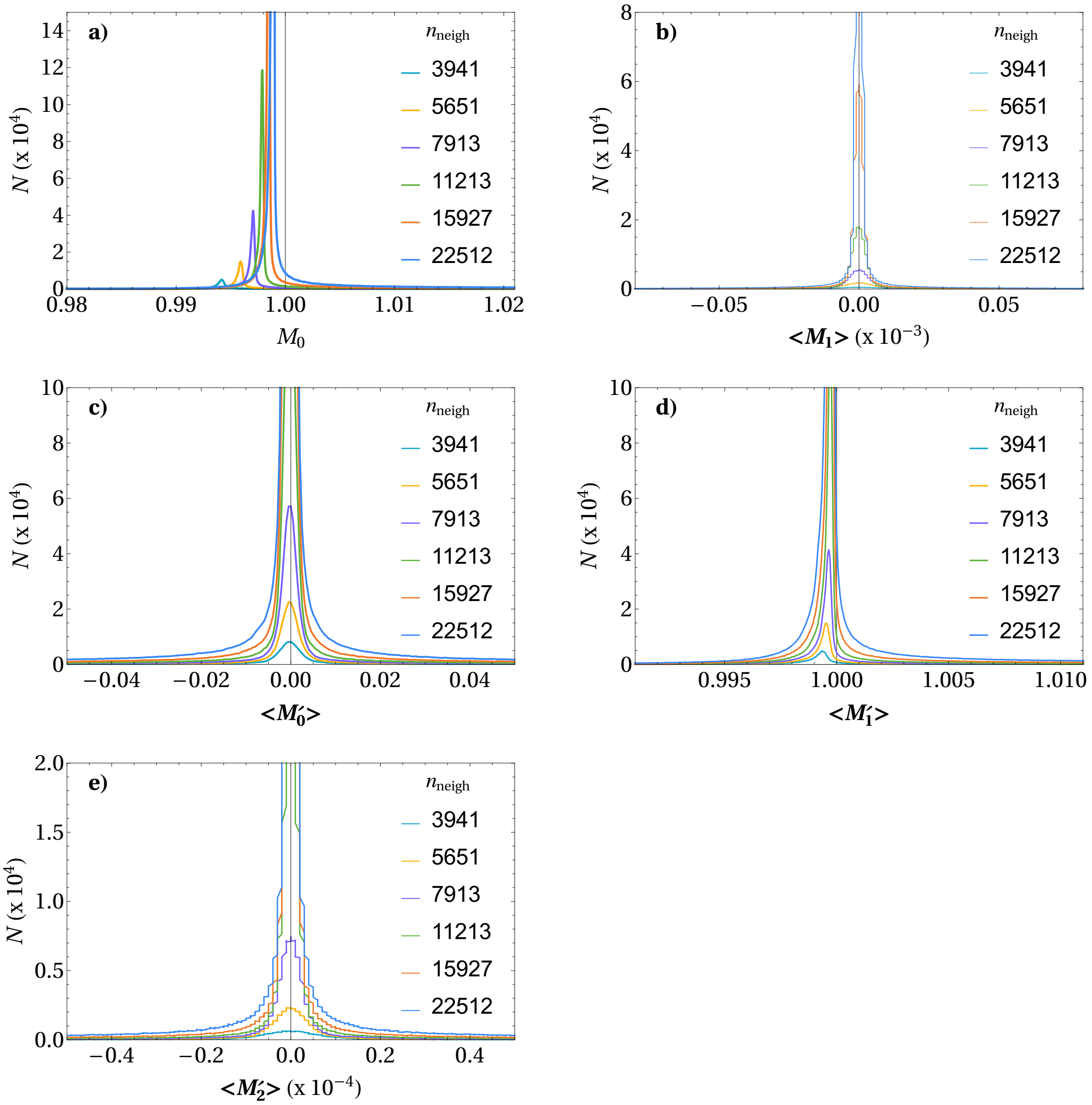

**Figure 13.** Distributions of the moments of the kernel $M_0$ and $\mathbf{M}_1$ as given by Eqs. (13) and (28) (upper plots) and the kernel gradient $\mathbf{M}'_0$, $\mathbf{M}'_1$, and $\mathbf{M}'_2$ as given by Eqs. (24), (14), and (30) (middle and bottom plots), respectively, for models B1W$_{\mathrm{ADIA}}$–B6W$_{\mathrm{ADIA}}$. $\langle \mathbf{M}_1 \rangle$ and $\langle \mathbf{M}'_0 \rangle$ are the average values of the components of vectors $\mathbf{M}_1$ and $\mathbf{M}'_0$, while $\langle \mathbf{M}'_1 \rangle$ and $\langle \mathbf{M}'_2 \rangle$ correspond to the average of the components of tensors $\mathbf{M}'_1$ and $\mathbf{M}'_2$, respectively. The vertical solid line in each frame marks the value of the kernel moments for which $C^1$-consistency is achieved.

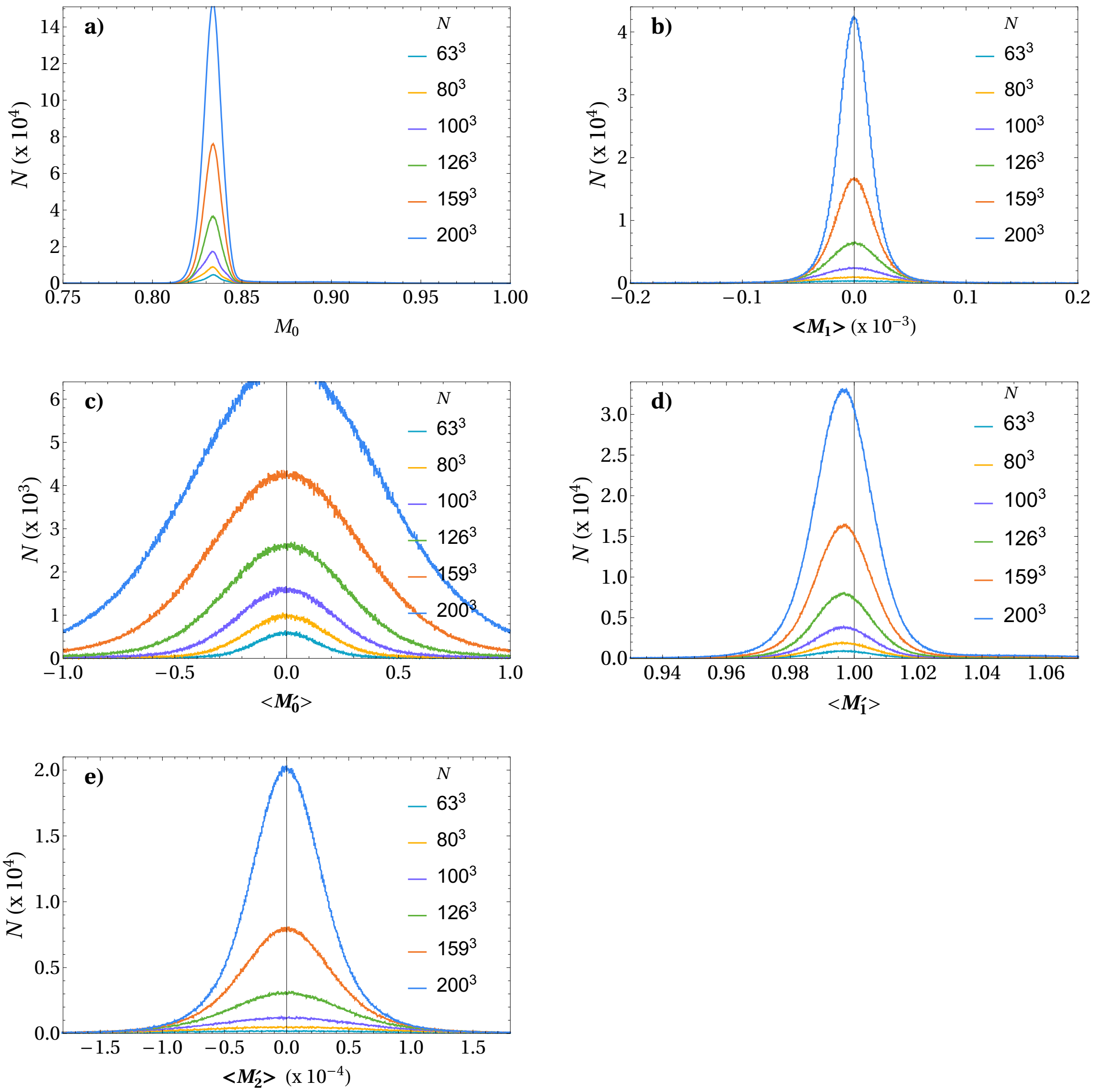

**Figure 14.** The same as Fig. 13 for models B1C$_{\rm ADIA}$–B6C$_{\rm ADIA}$.

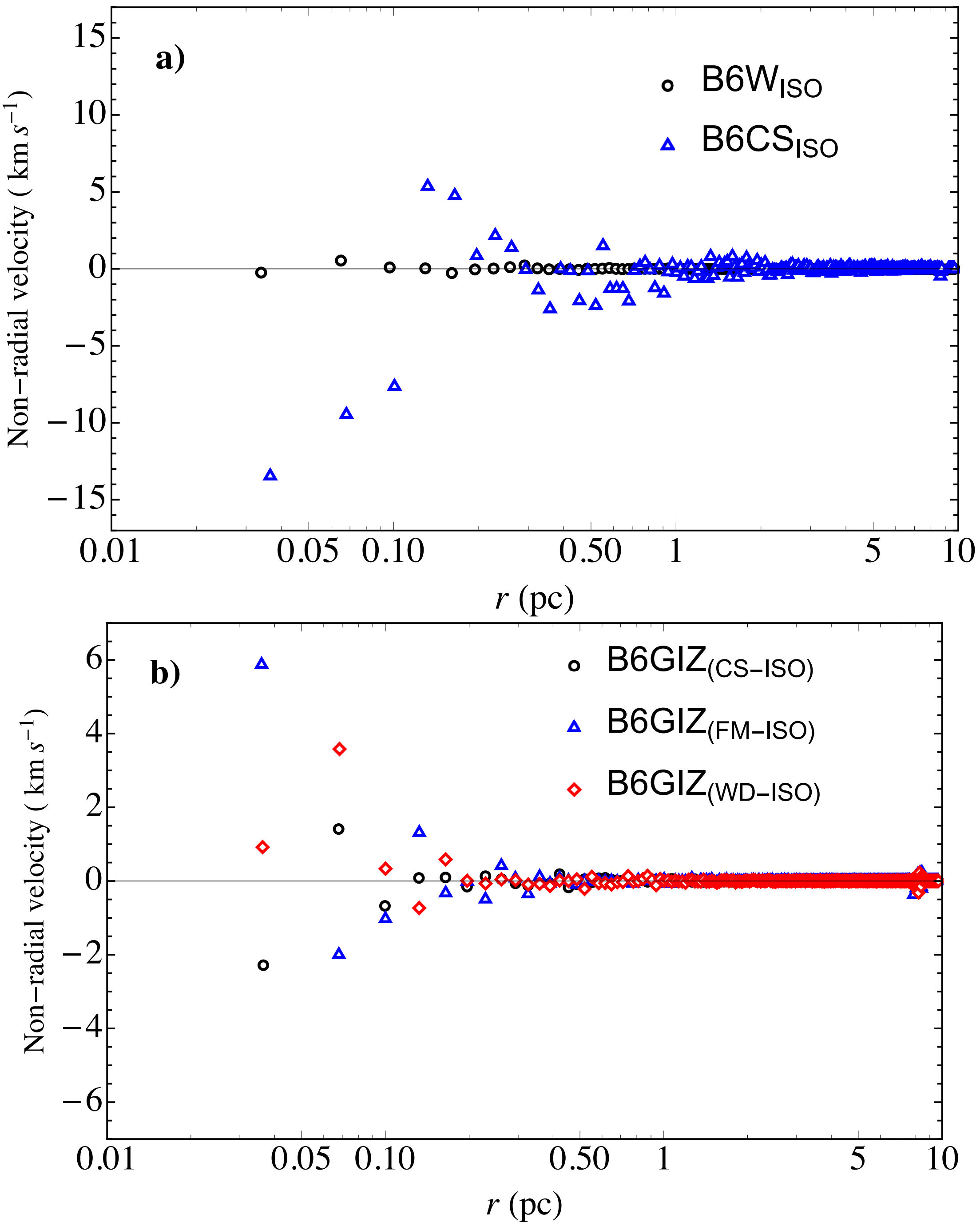

**Figure 15.** (a) Average values of the non-radial velocities taken over concentric shells of width $\approx 0.033$ pc as functions of radius for the more representative models B6C$_{\rm ISO}$ and B6W$_{\rm ISO}$ and (b) for the GIZMO simulations B6GIZ$_{\rm (CS-ISO)}$, B6GIZ$_{\rm (WD-ISO)}$, and IZ$_{\rm (FM-ISO)}$. The scatter in the data close to the accretor measures the deviations from spherical symmetry due to non-linear amplification of numerical noise.